\documentclass[]{aastex631}

\usepackage{comment}
\usepackage{hyperref}
\usepackage{amsmath}
\usepackage{cleveref}
\usepackage{enumitem}
\usepackage{graphicx}
\usepackage{lmodern}

\shorttitle{CHARA images of $\rho$~Cas}
\shortauthors{Anugu et al.}

\begin{document}

\title{CHARA Near-Infrared Imaging of the Yellow Hypergiant Star $\rho$~Cassiopeiae: \
Convection Cells and Circumstellar Envelope }

\correspondingauthor{Narsireddy Anugu} \email{nanugu@gsu.edu}

\author[0000-0002-2208-6541]{Narsireddy Anugu}
\affiliation{The CHARA Array of Georgia State University, Mount Wilson Observatory, Mount Wilson, CA 91023, USA}

\author[0000-0002-5074-1128]{Fabien Baron}
\affiliation{Center for High Angular Resolution Astronomy and Department 
of Physics and Astronomy, Georgia State University, P.O. Box 5060, Atlanta,
GA 30302-5060, USA}

\author[0000-0002-3380-3307]{John D. Monnier}
\affiliation{Department of Astronomy, University of Michigan, Ann Arbor, MI 48109, USA}

\author[0000-0001-8537-3583]{Douglas R. Gies}
\affiliation{Center for High Angular Resolution Astronomy and Department 
of Physics and Astronomy, Georgia State University, P.O. Box 5060, Atlanta,
GA 30302-5060, USA}

\author[0000-0002-9288-3482]{Rachael M. Roettenbacher}
\affiliation{Department of Astronomy, University of Michigan, Ann Arbor, MI 48109, USA}

\author[0000-0001-5415-9189]{Gail H. Schaefer}
\affiliation{The CHARA Array of Georgia State University, Mount Wilson Observatory, Mount Wilson, CA 91023, USA}

\author[0000-0002-7540-999X]{Miguel Montarg\`es}
\affiliation{LESIA, Observatoire de Paris, Universit\'e PSL, CNRS, Sorbonne Universit\'e, Université Paris Cit\'e, 5 place Jules Janssen, 92195 Meudon, France}

\author[0000-0001-6017-8773]{Stefan Kraus}
\affiliation{Astrophysics Group, Department of Physics \& Astronomy, University of Exeter, Stocker Road, Exeter, EX4 4QL, UK}

\author[0000-0002-0493-4674]{Jean-Baptiste Le Bouquin}
\affiliation{Institut de Planetologie et d\'Astrophysique de Grenoble, Grenoble 38058, France}

\author[0000-0002-9759-038X]{Matthew D. Anderson}
\affiliation{Georgia Tech Research Institute, Baker Building, 925 Dalney St. Atlanta, GA 30318}

\author[0000-0002-0114-7915]{Theo ten Brummelaar}
\affiliation{The CHARA Array of Georgia State University, Mount Wilson Observatory, Mount Wilson, CA 91023, USA}

\author{Isabelle Codron}
\affiliation{Astrophysics Group, Department of Physics \& Astronomy, University of Exeter, Stocker Road, Exeter, EX4 4QL, UK}

\author[0000-0001-9939-2830]{Christopher D. Farrington}
\affiliation{The CHARA Array of Georgia State University, Mount Wilson Observatory, Mount Wilson, CA 91023, USA}

\author[0000-0002-3003-3183]{Tyler Gardner}
\affiliation{Astrophysics Group, Department of Physics \& Astronomy, University of Exeter, Stocker Road, Exeter, EX4 4QL, UK}

\author{Mayra Gutierrez}
\affiliation{Department of Astronomy, University of Michigan, Ann Arbor, MI 48109, USA}

\author{Rainer K\"{o}hler}
\affiliation{The CHARA Array of Georgia State University, Mount Wilson Observatory, Mount Wilson, CA 91023, USA}

\author[0000-0001-9745-5834]{Cyprien Lanthermann}
\affiliation{The CHARA Array of Georgia State University, Mount Wilson Observatory, Mount Wilson, CA 91023, USA}

\author[0000-0002-9120-9728]{Ryan Norris}
\affiliation{New Mexico Institute of Mining and Technology, Workman Center, 801 Leroy Place, Socorro, NM 87801, USA}

\author[0000-0003-1038-9702]{Nicholas J. Scott}
\affiliation{The CHARA Array of Georgia State University, Mount Wilson Observatory, Mount Wilson, CA 91023, USA}

\author[0000-0001-5980-0246]{Benjamin R. Setterholm}
\affiliation{Department of Astronomy, University of Michigan, Ann Arbor, MI 48109, USA}

\author{Norman L. Vargas}
\affiliation{The CHARA Array of Georgia State University, Mount Wilson Observatory, Mount Wilson, CA 91023, USA}

\begin{abstract} 
Massive evolved stars such as red supergiants and hypergiants are potential progenitors of Type II supernovae, and they are known for ejecting substantial amounts of matter,  up to half their initial mass, during their final evolutionary phases. The rate and mechanism of this mass loss play a crucial role in determining their ultimate fate and the likelihood of their progression to supernovae. However, the exact mechanisms driving this mass ejection have long been a subject of research. Recent observations, such as the Great Dimming of Betelgeuse, have suggested that the activity of large convective cells, combined with pulsation, could be a plausible explanation for such mass loss events.
In this context, we conducted interferometric observations of the famous yellow hypergiant, $\rho$~Cassiopeiae  using the CHARA Array in H and K-band wavelengths. $\rho$~Cas is well known for its recurrent eruptions, characterized by periods of visual dimming ($\sim$~1.5-2 mag) followed by recovery. From our observations, we derived the diameter of the limb-darkened disk and found that this star has a radius of $1.04\pm0.01$ milliarcseconds (mas), or $564 - 700~R_\odot$.
We performed image reconstructions with three different image reconstruction software packages, and they unveiled the presence of giant hot and cold spots on the stellar surface. We interpret these prominent hot spots as giant convection cells, suggesting a possible connection to mass ejections from the star's envelope. Furthermore, we detected spectral CO emission lines in the K-band ($\lambda=2.31-2.38 ~\mu$m), and the image reconstructions in these spectral lines revealed an extended circumstellar envelope with a radius of $1.45\pm0.10$ mas.
\end{abstract}


\section{Introduction}
Yellow hypergiants (YHGs) represent a brief yet pivotal evolutionary phase for massive stars as they transition from or to red supergiant phases \citep{deJager1998A&ARv...8..145D}. These YHGs typically have initial masses of $20-40 M_\odot$ \citep{Ekstrom2012A&A...537A.146E, Kourniotis2022MNRAS.511.4360K} and exhibit high luminosity ranging from $5.4 \leq \log(L/L_\odot) \leq 5.8$ \citep{Neugent2012ApJ...749..177N, Dorn-Wallenstein2023ApJ...959..102D, Humphreys2023AJ....166...50H}. Their effective temperatures fall approximately within the range of $T_{\rm eff}\sim4000-7500$ K \citep[excluding eruptive episodes, ][]{Neugent2012ApJ...749..177N}. One of their defining characteristics is their remarkable variability in spectral type in a short period of time.  This spectral variability is linked to another unique feature: their inherent instability around a temperature of $T_{\rm eff}\sim7000-12000$~K. In this temperature region, YHGs experience an instability zone known as the ``Yellow Void" \citep{Nieuwenhuijzen1995A&A...302..811N, deJager1998A&ARv...8..145D, Nieuwenhuijzen2000A&A...353..163N, Stothers2001ApJ...560..934S, Glatzel2024MNRAS.529.4947G}. In this yellow void region, the effective gravity force becomes zero, leading to episodic mass loss in these stars. As the star loses mass, its effective temperature decreases, causing it to move redwards on the H-R diagram and leave the yellow void. Subsequently, the temperature starts to increase, moving the star move bluewards again on the H-R diagram until it enters the yellow void once more. The star may oscillate between hot and cool evolutionary phases,  undergoing multiple ``blue loops" \citep{deJager1998A&ARv...8..145D, Stothers2012}.  For instance, $\rho$~Cas lost matter at a rate of up to $5.4\times10^{-2}M_\odot{\rm yr}^{-1}$ during the mass loss episode it experienced in the years 2000-2001 \citep{Lobel2003ApJ...583..923L}.  These mass loss episodes often manifest as visual dimming events followed by subsequent brightening in the star's light curve. The dimming occurs because newly ejected material condenses into dust around the star, temporarily obscuring it from our view. This phenomenon is similar to the dimming events observed in other massive stars like Betelgeuse \citep{Montarges2021} and RW Cep  \citep{Anugu2023}.  Spectroscopic observations during these dimming events reveal a rapid drop in temperature and the presence of expanding circumstellar shells. These shells are detected by analyzing specific spectral lines, such as CO band emissions and the wings of the H$\alpha$ line \citep{Lobel2003ApJ...583..923L}.

The significant mass loss  experienced during the YHG phase profoundly influences the star's fate, ultimately determining its position on the H-R diagram at the end of its life \citep{deJager1998A&ARv...8..145D}. This mass loss can lead to the transformation of the YHG stars into hot supergiants such as luminous blue variables, B[e] supergiants, or Wolf–Rayet stars, as well as influence the type of supernova and resulting remnant. Furthermore, a few potential YHG supernova progenitors have been identified, but controversy surrounds them because they do not fully meet the mass and luminosity criteria typically associated with YHGs \citep{Maund2011ApJ...739L..37M,Georgy2012A&A...538L...8G, Groh2014A&A...572L..11G, Mauerhan2015MNRAS.447.1922M,Klochkova2019AstBu..74..475K, Kilpatrick2021MNRAS.504.2073K}.

 YHGs serve as invaluable laboratories for investigating the mechanisms behind mass loss events and for testing theories of stellar evolution, given the active and pivotal role of the YHG phase in post-red supergiant evolution. However, their rarity presents a significant challenge. They are relatively scarce because of their short-lived phase, typically lasting less than $10^4$ years. Additionally, their substantial mass loss can lead to self-obscuration by dust formed from the cooled ejecta, which makes them difficult to detect directly. Currently, we are aware of approximately 35 objects classified as YHGs or YHG candidates in our Milky Way Galaxy \citep{deJager1998A&ARv...8..145D} and neighboring galaxies of the Local Group, such as the Small Magellanic Cloud \citep{Neugent2010ApJ...719.1784N,Kourniotis2022MNRAS.511.4360K}, the Large Magellanic Cloud \citep{Neugent2012ApJ...749..177N,Humphreys2023AJ....166...50H,Dorn-Wallenstein2023ApJ...959..102D}, and M33 \citep{Humphreys2006AJ....131.2105H,Drout2012ApJ...750...97D}. Consequently, only a handful have been extensively studied, with IRC+10420 and $\rho$\,Cas being two of the most prominent examples. Both these stars have asymmetric circumstellar shells and have undergone episodic, high-mass loss events spanning more than a hundred years. IRC+10420, with $T_{\rm eff}=7930\pm140$~K \citep{Nieuwenhuijzen2000A&A...353..163N} has been seen to be evolving in the H-R diagram from the red to the blue and is believed to be crossing the Yellow Void to evolve as a low-luminosity Luminous Blue Variable. 
 This dramatic shift is evidenced by a change in spectral type from F8Ia to early A5Ia in 30 yr, indicating a rapid temperature increase of $\sim120$ K per year \citep{Oudmaijer1996MNRAS.280.1062O, Oudmaijer1998A&AS..129..541O, Klochkova1997MNRAS.292...19K, Klochkova2002ARep...46..139K}.  Additionally, it experienced a mass loss rate of $2\times10^{-3}M_\odot{\rm yr}^{-1}$ until about 2000 years ago  and since then the rate decreased \citep{Shenoy2016}. This expelled material has sculpted a complex circumstellar environment, filled with knots, arcs, shells, and even jet-like structures \citep{Tiffany2010AJ....140..339T}.

\begin{figure}[h]
\centering
\includegraphics[width=\textwidth]{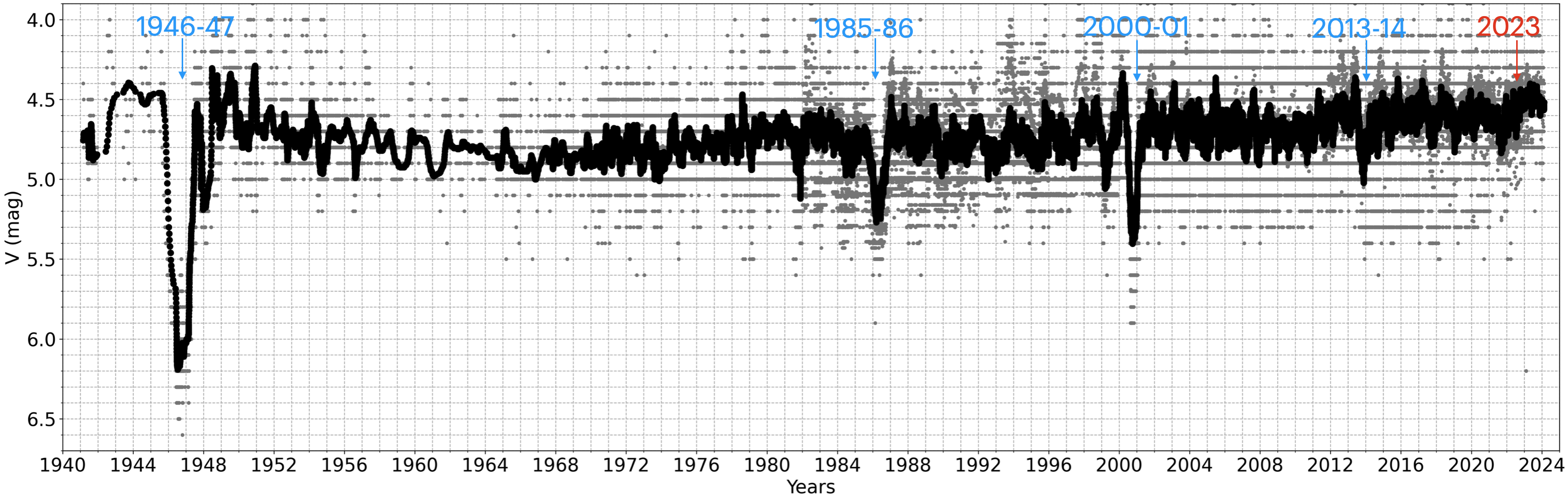}
\caption{The apparent visual brightness of $\rho$~Cas over the period from 1942 to 2024. The strongest visual dimming was recorded in 1945-47 with much longer duration compared to others. The grey and black points represent the raw and 10-day averaged and median-filtered data from the American Association for Variable Star Observers (AAVSO), respectively. The great dimming event in 1946 stands out when compared to other dimming occurrences in 1986, 2000, and 2013. During our CHARA observation epochs in Oct/Nov 2023 (red color), there was no notable dimming activity.   }
\label{Fig:light_curve}
\end{figure}

In this study, we focus on $\rho$~Cas, an archetype YHG. Table~\ref{Table:properties} summarizes the properties of $\rho$~Cas, confirming its status as a YHG.   With a mass of $\sim40M_\odot$, luminosity range $\log(L/L_\odot) \sim 5.48 - 5.72$, and effective temperature averaging $\sim7000$ K, $\rho$~Cas aligns perfectly with the established parameters for YHGs. Additionally, $\rho$~Cas is a well-known for its historical outburst episodes of dimming and re-brightening, with a visual brightness change of approximately $1.5-2$ magnitudes (see Figure~\ref{Fig:light_curve}). It experienced four documented outbursts in the last century that happened in  1945-1947 \citep{Beardsley1961ApJS....5..381B}, 1985-1986 \citep{Boyarchuk1988TarOT..92...40B,Leiker1989IBVS.3345....1L}, 2000-2001 \citep[the `Millenium outburst’, ][]{Lobel2003ApJ...583..923L}, and 2013-2014 \citep{Kraus2019MNRAS.483.3792K}. Additionally, $\rho$~Cas is known for a dramatic decrease in the effective temperature by 3000~K due to the ejection of a hydrogen shell \citep{Lobel2003ApJ...583..923L}.  These outbursts in $\rho$~Cas resulted in the formation of circumstellar material and multiple detached shells, which were observed using spectroscopic  $H\alpha$ wings, Ca II,  and CO  band emissions \citep{Sargent1961ApJ...134..142S, Lambert1981ApJ...248..638L, Lobel1998, Lobel2003ApJ...583..923L, Kraus2019MNRAS.483.3792K}. A high-spectral resolution analysis by \cite{Lobel2003ApJ...583..923L} reported that during the 2000-2001 outburst, the mass loss in this star reached a peak of approximately $5.4\times10^{-2}M_\odot{\rm yr}^{-1}$,  resulting in the formation of an optically thick circumstellar gas shell
of $3\times10^{-2}M_\odot$ during 200 days after outburst.  
The analysis of the visual light curve reveals that the outburst episodes are not only becoming more frequent, but also exhibiting a reduction in their duration \citep{Maravelias2022JAVSO..50...49M}. This observed pattern hints at the possibility that $\rho$~Cas may be transitioning into the next phase of its evolutionary cycle traversing the Yellow Void \citep[][references within]{Kraus2019MNRAS.483.3792K}. 
Analysis of infrared photometry data suggests dust formation occurred between 1973 and 1983, likely from material ejected during the powerful 1946 outburst \citep{Jura1990ApJ...351..583J}. However recent mid-infrared observations find no evidence of subsequent dust formation after 1946 \citep{Shenoy2016}.  This suggests that $\rho$~Cas might be in an earlier evolutionary stage compared to other hypergiants (such as IRC+10420  and HD~179821) that show a significant infrared excess from dust shells \citep{Castro-Carrizo2007A&A...465..457C, Quintana-Lacaci2008A&A...488..203Q}.

In this paper, our objective is to study the fundamental parameters of $\rho$~Cas to gain insight into the mechanisms driving its mass loss. The main mechanisms that contribute to mass loss are pulsational-driven and radiation pressure line-driven stellar winds \citep{Nathan2014ARA&A..52..487S}. The study of the Great Dimming episode of Betelgeuse suggests that the trigger for mass loss may be related to the combined effects of giant convective cells on the stellar surface, pulsations, and the magnetic field of the star \citep{Humphreys2022, Dupree2022}. However, distinguishing between these mechanisms remains challenging. Our goal is to enhance our understanding of these underlying mechanisms by imaging the stellar surface and circumstellar environment of $\rho$~Cas.

Here, we present the first high-angular-resolution images of $\rho$~Cas  photosphere, captured using the Georgia State University Center for High Angular Resolution Astronomy (CHARA) Array situated at the Mount Wilson Observatory \citep{tenBrummelaar2005, Schaefer2020}. The CHARA images reveal unprecedented details of the stellar surface, including giant convective cells. These cells are detected as hot and cool spots on the stellar surface, indicative of stellar activity,  which may have triggered the episodic mass loss events observed in the past. We describe the interferometric observations and the reduction of the data from the CHARA Array in Section~\ref{Sec:obser}. As a first step in the data analysis, we derived diameter measurements using uniform disk (UD) and limb-darkened disk (LDD) model fitting (Section~\ref{Sec:ldd_fit}). We utilized three different image reconstruction techniques to extract images from the interferometric observations. The image reconstructions of the continuum and carbon monoxide (CO) spectral line bands are provided in Section~\ref{Sec:images} and Section~\ref{Sec:CO-lines}, respectively. Finally, we compare $\rho$~Cas with other hypergiants and discuss the implications of these observations for models of dimming and mass ejection in Section~\ref{Sec:discussion}.

\begin{deluxetable*}{l l l l l l  }
\tablecaption{$\rho$~Cas  properties.}
\label{Table:properties}
\tablewidth{0pt}
\tablehead{
\colhead{Parameter} &  \colhead{Value} &  \colhead{Reference}
}
\startdata
Stellar photosphere radius (mas) & $1.05 \pm 0.01$ & This work \\
Circumstellar env radius (mas) & $1.45 \pm 0.01$ & This work \\
Stellar photosphere radius ($R_\odot$) & $564 - 700$ & This work\\
Circumstellar env radius ($R_\odot$) & $780 - 970$ & This work\\
Spectral type & F8-G2 Ia0 & \citet{Klochkova2014}\\
             &  & \citet{Sblewski2022} \\
Distance (kpc) & 2.5 - 3.1   & \citet{vanGenderen2019} \\
& & \citet{Gorlova2006} \\
Temperature range (K) & 5000 - 7250 & \citet{Lobel1998} \\ 
Mean temperature (K) & $\sim7000$& \citet{Kraus2019MNRAS.483.3792K}\\
& & \citet{vanGenderen2019} \\
Luminosity ($L/L_\odot\times10^3$) & $302-530$ & \citet{vanGenderen2019} \\
Stellar mass ($M/M_\odot$) & $\sim40$  & \citet{Gorlova2006} \\
Age (Myr) & 4 - 6 & \citet{Gorlova2006} \\
Short primary period (d) & 380-650 &  \citet{Percy2000}\\
Long secondary period (d) & 820 &  \citet{Percy2000}\\
Projected rotational velocity (km s$^{-1}$) &	25 &\citet{Lobel1998}
\enddata
\end{deluxetable*}

\begin{deluxetable*}{l l l l l l  }
\tablecaption{$\rho$~Cas and calibrator observing log. The calibrator star's H-band uniform disk diameters are adopted from the JMMC catalog~\citep{Bourges2017}. The difference in H- and K-band diameters is less than 1\%, and it is smaller than their diameter uncertainty.}
\label{Table:ObsLog}
\tablewidth{0pt}
\tablehead{
\colhead{UT date} &  Calibrators & Diameter  & Seeing Fried Param\\
\colhead{} &   & (mas) & $r_0$ (cm)
}
\startdata
2023 Oct 25 & HD 6238 & $0.55 \pm 0.01$   & 7.45 \\
           & HD 219080 & $0.69 \pm 0.07$ & \\
           & HD 211982 & $0.57 \pm 0.02$ & \\
2023 Oct 28 & HD 6238 & $0.55 \pm 0.01$  & 5.2 \\
           & HD 219080 & $0.69 \pm 0.07$ & \\
           & HD 13137 & $0.66\pm0.04$ & \\
           & HD 554 & $0.58\pm0.01$ & \\
2023 Oct 29 & HD 6238 & $0.55 \pm 0.01$ & 7.4 \\
           & HD 219080 & $0.69 \pm 0.07$ & \\
2023 Nov 14 & HD 6027 & $0.73 \pm 0.02$ & 9.5 \\
           & HD 6238 & $0.55 \pm 0.01$ & \\
           & HD 219080 & $0.69 \pm 0.07$ & \\
           & HD 13137 & $0.66\pm0.04$ & \\
\enddata
\end{deluxetable*}

\begin{figure}[h!]
\includegraphics[width=0.5\textwidth]{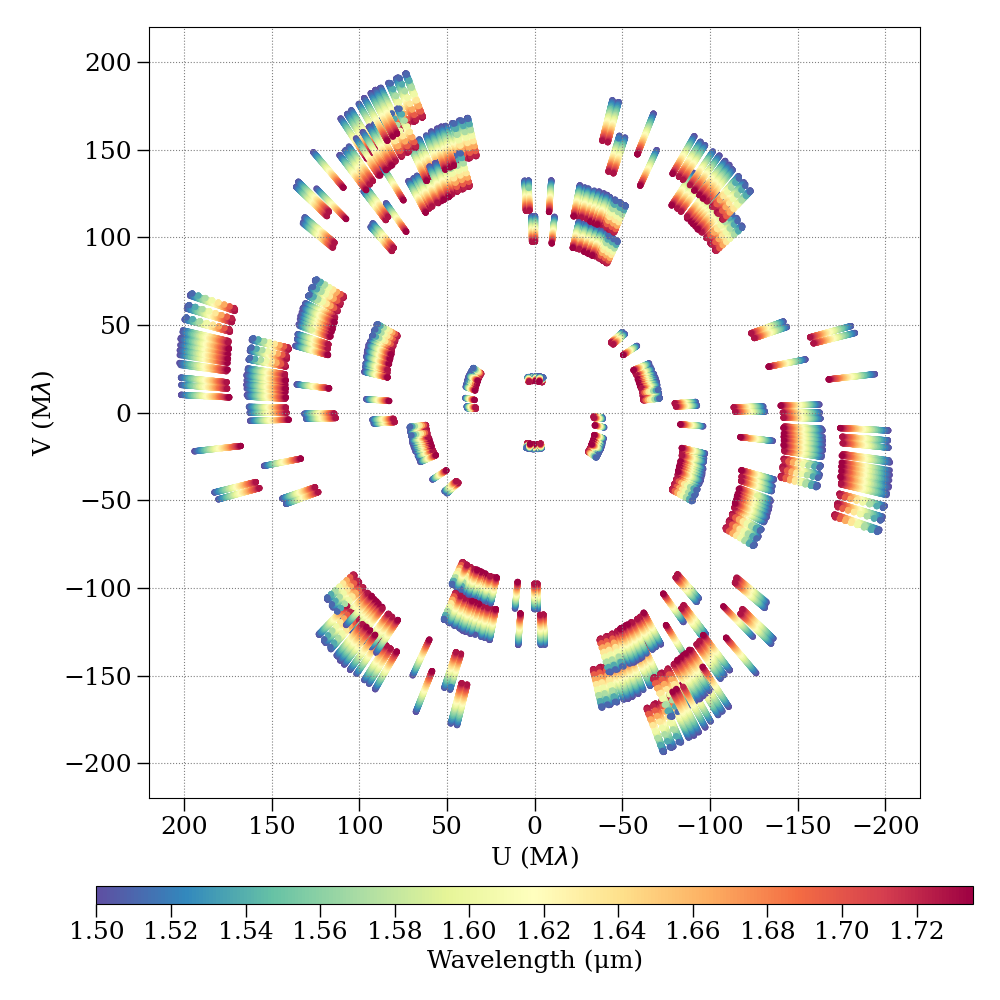}
\includegraphics[width=0.5\textwidth]{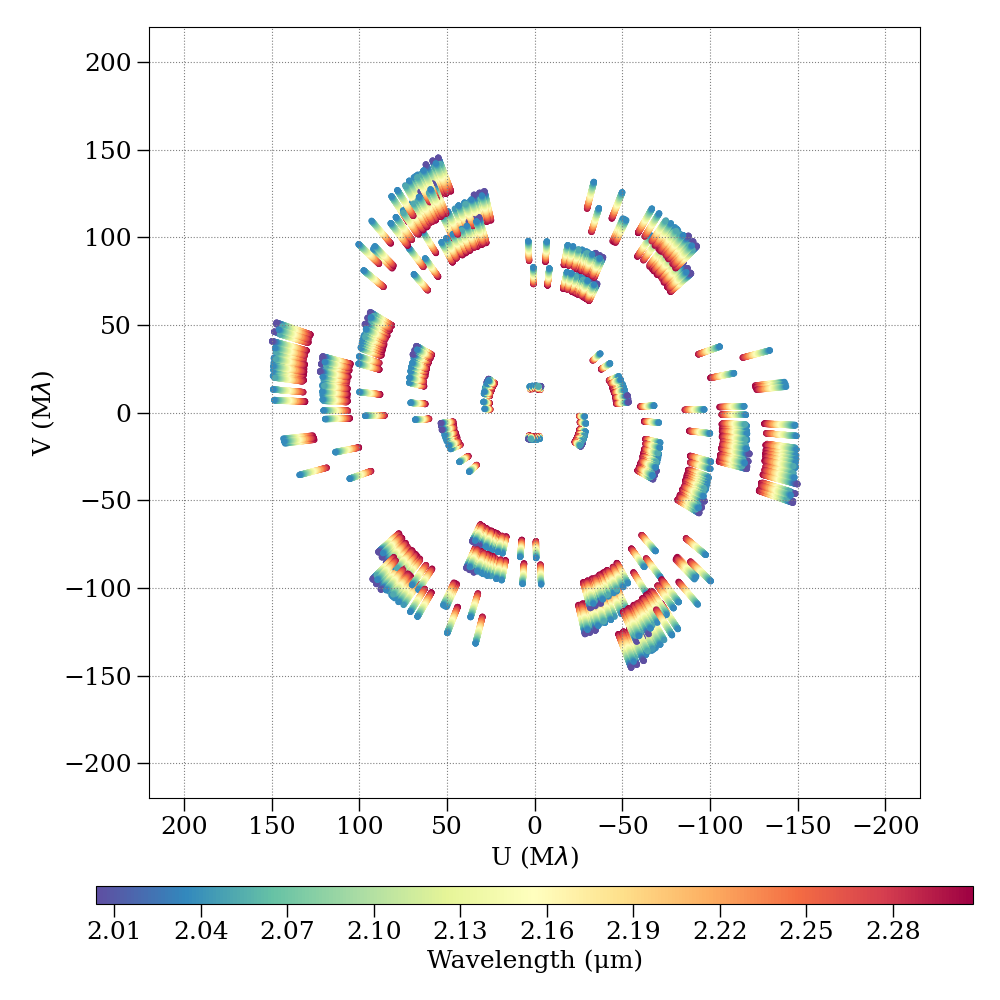}
\caption{The spatial frequency coverage in the $(u,v)$ plane for the 
$H$-band (left) and $K$-band (right) observations. The  color 
legends show the corresponding wavelengths.  }
\label{fig_uv}
\end{figure}

\begin{figure}[h!]
\centering
\includegraphics[width=\textwidth]{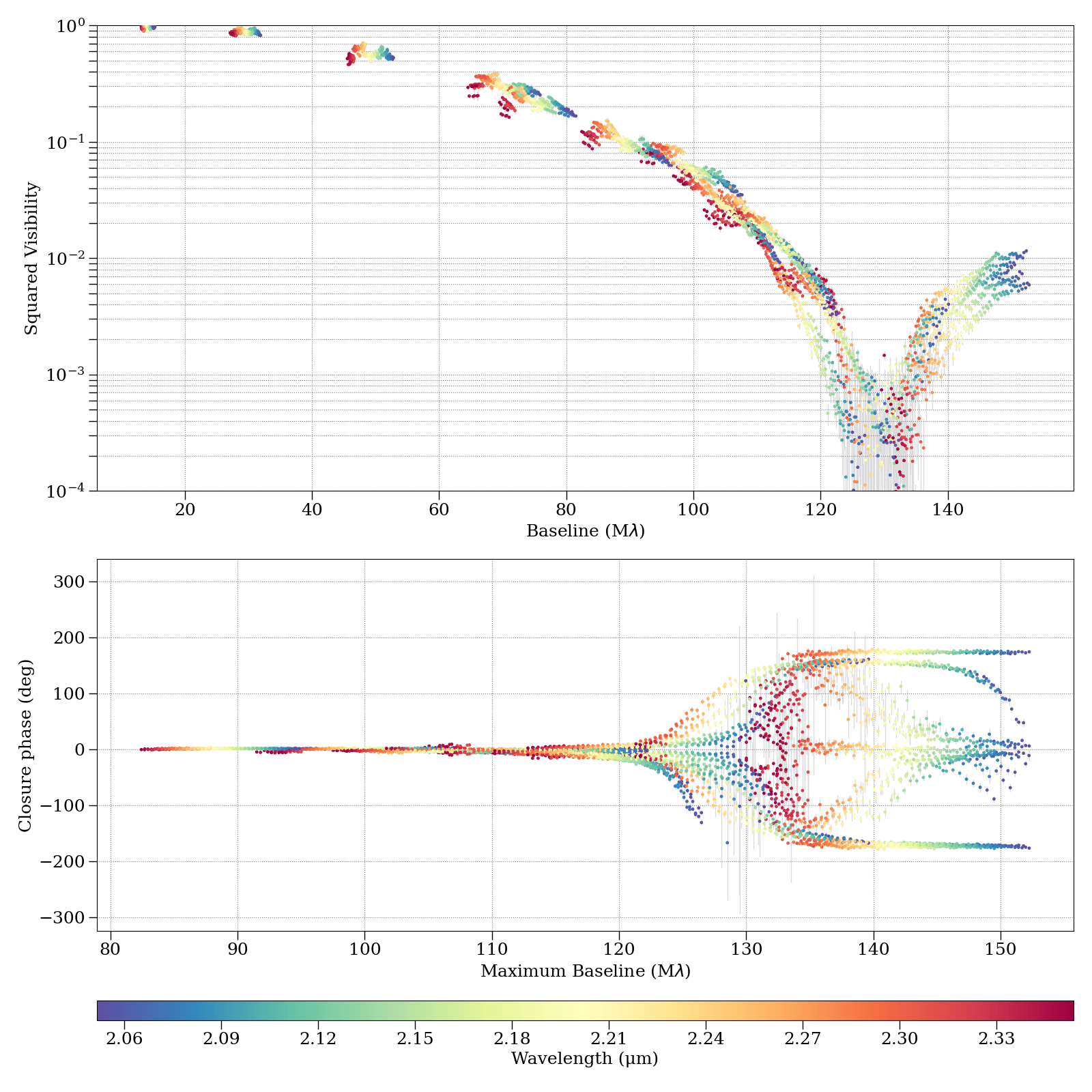}
\caption{  
CHARA/MYSTIC calibrated observations of $\rho$~Cas observed on 2023 Oct 25. Top panel: squared visibility, and bottom panel: closure phases.  The color bar indicates the wavelength. The $V^2$ is a measure of whether the object is spatially resolved, i.e., decreasing $V^2$ with spatial frequency ($B/\lambda$) indicates that the observed object is spatially resolved. Non-zero closure phases indicate that the source is not point-symmetric, which is typically caused by binary and star surface features. 
}
\label{Fig:v2_t3phi}
\end{figure}

\section{CHARA Observations}\label{Sec:obser}

We observed $\rho$~Cas at four epochs using the CHARA Array between 2023 October and November (see Table~\ref{Table:ObsLog}). The CHARA Array is the world's largest optical and near-infrared interferometer, having baseline lengths spanning from $B=34$~m to 331~m.  In the H-band ($\lambda=1.50 - 1.74 ~\mu$m) and K-band ($2.00 - 2.38 ~\mu$m) wavelength range, it provides angular resolutions of approximately $\lambda/2B \sim 0.5$ and 0.6 milliarcseconds (mas), respectively. We used the H-band  MIRC-X \citep{Anugu2020} and K-band MYSTIC \citep{Setterholm2022} beam combiners in a cophasing observing mode. In this mode, we collected data simultaneously from both MIRC-X and MYSTIC instruments, with MYSTIC serving as the primary fringe tracker, controlling the CHARA main delay lines, while MIRC-X functioned as the secondary science beam combiner, managing internal differential delay lines. MIRC-X and MYSTIC are six telescope beam combiners and deliver 15 squared visibilities ($V^2$) and 10  independent closure phases ($\phi$) in each operating wavelength band.  The observations were carried out with a spectral resolving power of $R=190$ for MIRC-X and $R=278$ for MYSTIC. The observations cover 30 and 52 spectral channels across the near-infrared H and K-bands.

To ensure better calibration of our observations, we employed multiple calibrators (see Table \ref{Table:ObsLog}). We chose calibrators by cross-checking with the Gaia archive to avoid known binaries. These calibrators are used to correct any visibility loss in the target observations caused by time-variable factors, including fluctuations in atmospheric coherence time, vibrations, differential dispersion, and birefringence in the beam path.

The raw data sets are available on the CHARA data archive portal\footnote{\href{https://www.chara.gsu.edu/observers/database}{https://www.chara.gsu.edu/observers/database}} and these raw datasets are reduced using the standard \texttt{mircx pipeline} \citep{ Anugu2020,le_bouquin_2024_12735292}. The pipeline provides uncalibrated oifits files of the science target comprising measurements of the spectral distributions of flux, squared visibility,  and closure phases. To assess whether the calibrator stars displayed any binary signatures, we calibrated the data from these calibrator stars against each other. To detect binary signatures in these data, we fit models using a grid-search script from the Python-based Parametric Modeling of Optical InteRferomEtric Data software, known as PMOIRED \citep{antoine_merand_2024_10889235, Merand2022}.  We found that the calibrator stars did not exhibit any binary signatures. The calibration was achieved with the IDL mircx\_cal.script \citep{Monnier2012ApJ...761L...3M,Anugu2020}. 

On the night of 2023 October 28, there was a relatively higher systematic visibility bias attributed to atmospheric seeing conditions. During our investigation, we performed imaging analyses both with and without incorporating the data from this particular night. The results did not exhibit substantial differences that might impact the overall outcomes of our study. Therefore, we opted to exclude the 2023 October 28 dataset in this paper.

Figure~\ref{fig_uv} illustrates the $(u,v)$ plane coverage of the combined observations, which provides a good uniform sampling and yields high-quality image reconstructions. The effective resolution of interferometric synthetic images is fundamentally determined by the maximum extent of the $(u,v)$-coverage.  The sparseness or incompleteness in the filling of the $(u,v)$-plane tends to create  artifacts in these images.  Figure~\ref{Fig:v2_t3phi} shows the visibilities $V^2$ and closure phases ($\phi$) of $\rho$~Cas in the K band. The closure phases exhibit large, non-zero values that vary significantly with wavelength and show markedly different behaviors among the various triplets of telescopes. This indicates that the target has a strongly asymmetric spatial flux distribution.

\section{Stellar diameter}\label{Sec:ldd_fit}

\begin{figure}[h]
\centering
\includegraphics[width=\textwidth]{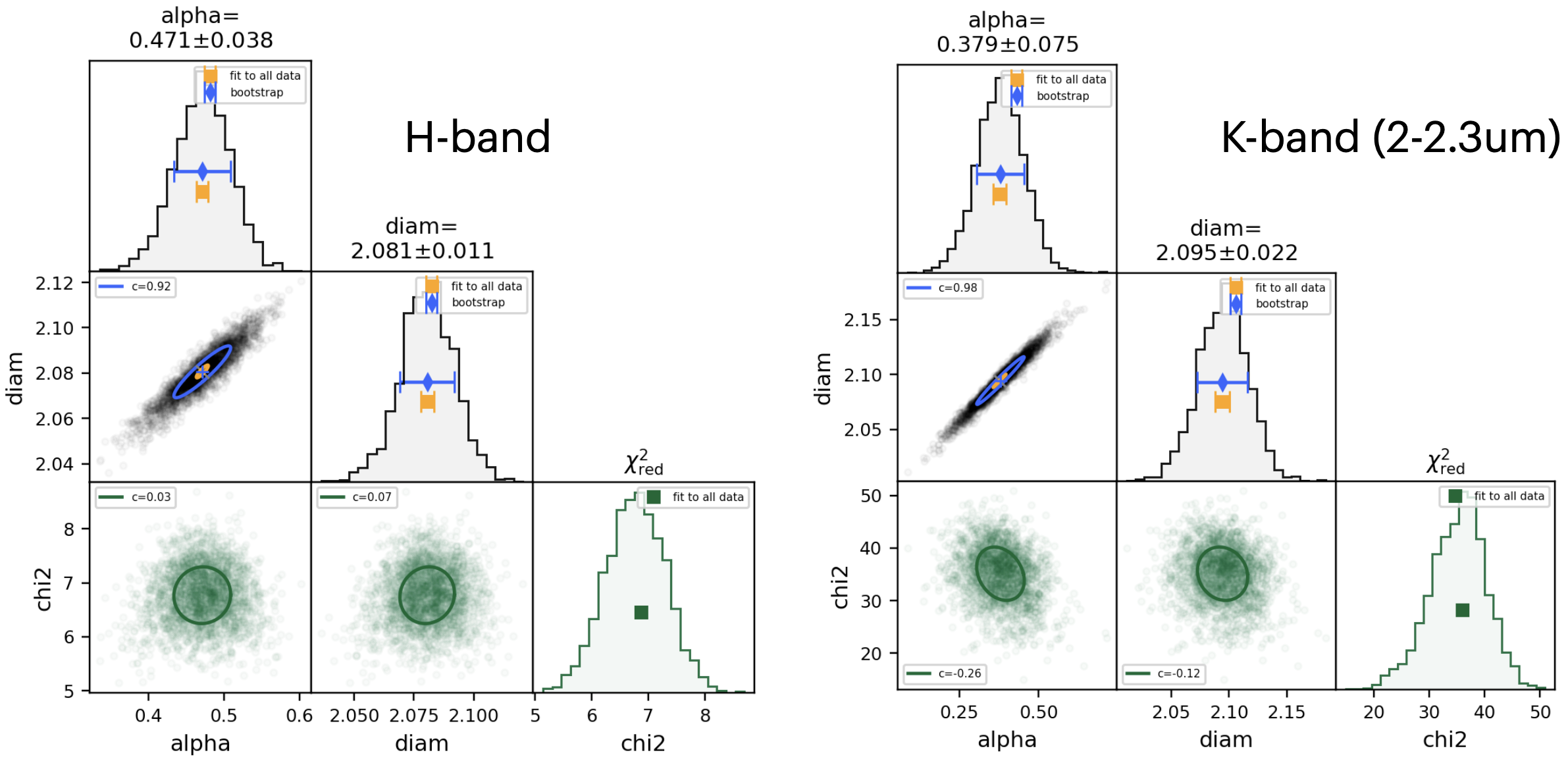}
\caption{Corner plot for the limb-darkened disk fit of diameter and power-law coefficient $\alpha$. The left and right panels correspond to the H- and K-band wavelengths, respectively. Uncertainties are estimated through bootstrap fitting of 500 times on the data to assess the scatter of the fitted parameters. The bootstrap results are plotted with $\sigma=4.5$ clipping. 
}
\label{Fig:oitools_LDD_fit}
\end{figure}

We initially focused on determining the angular diameter of the star from the continuum dataset. To achieve this, we employed two distinct stellar disk models. The first model is the uniform disk model (UD), while the second is a power law limb-darkened disk model (LDD), $I(\mu)/I_0=\mu^\alpha$,  where $\mu$ is the cosine of the angle between the surface normal and line of sight and it is $0\leq\mu \leq1$. The power law coefficient is  $\alpha$, which varies $\alpha \geq 0$. We combined all the epochs into a single dataset and used PMOIRED software for this model fitting. The best-fit parameters are summarized in Table \ref{Table:LDD_fit}. Applying the uniform disk model, we calculated the star diameter to be $2.0\pm0.1$ mas. On the other hand, using the LDD model, we obtained a slightly different diameter of $2.1^{+0.01}_{-0.02}$ mas, with the limb-darkening coefficient $\alpha_{\rm H}=0.47\pm0.04$ at H-band and $\alpha_{\rm K}=0.38\pm0.08$ at K-band.  It is important to note that both models yielded relatively high fitting errors, as indicated by the reduced chi-squared $\chi^2> 7$ (see Figure~\ref{Fig:oitools_LDD_fit}). We also fit LDD  for the individual epochs and found that the diameter and $\alpha$ changed by $\pm0.1$.  The reason for the error scatter occurring between epochs is that this simple 2-parameter model is inadequate, as the star exhibits asymmetry in its shape (see Appendix, Section~\ref{Sec:limb_darkening_analysis}).  This limb-darkening fit overestimates $\alpha$ in the presence of surface features. See the estimations from the image reconstructions in Section~\ref{Sec:images}.

Figure~\ref{Fig:v2_t3phi} illustrates that the $V^2$ in the spectral channels covering the CO bandhead emission line, i.e., between wavelengths in the range of $\lambda = 2.3 - 2.38~\mu$m is lower than the continuum, suggesting that the target is more resolved as compared to the continuum. We measured the diameter of the molecular features of the CO first-overtone bands, and it is $2.91\pm0.19$ mas. The extended structure observed above the photosphere (with a diameter of $2.08^{+0.01}_{-0.02}$ mas) is interpreted as a circumstellar CO shell environment. It is also confirmed in the image reconstruction (see Section~\ref{Sec:CO-lines}). The presence of a circumstellar CO shell is a common phenomenon in massive evolved stars \citep[e.g.,][]{Perrin2005, Ohnaka2011, Montarges2014, Anugu2023}. 
The limb darkened angular diameter measurements are summarized in Table 3.  We expect that the star appears larger at 
wavelengths where the opacity is larger, and variations in the reported diameters from the visible to near-IR to CO-band 
appear to be approximately consistent with opacity variations.

We also investigated whether $\rho$ Cas might be a compact binary system considering its high $v\sin i = 25$ km/s (Table~\ref{Table:properties}), such as found for the yellow hypergiant HR~5171~A (HD~119796) that has been identified as a contact binary through interferometric observations \citep{Chesneau2014A&A...563A..71C}. To explore this possibility, we conducted a search for a companion using the PMOIRED grid search method within a field of view up to 50 mas. However, no signals indicative of the presence of a companion were detected. Long-term spectroscopic monitoring \citep[8.5~yr,][]{Lobel2003ApJ...583..923L} and photometric observations \citep{Maravelias2022JAVSO..50...49M} also support the conclusion that there is no companion associated with $\rho$~Cas.

\begin{deluxetable*}{ccccccccc}[h]
\tablecaption{LDD Diameter estimation \label{Table:LDD_fit} }
\tablewidth{0pt}
\tablehead{
\colhead{Wavelength} & 
\colhead{Diameter} & 
\colhead{Reference} &
\\ 
\colhead{($\mu$m)}  & 
\colhead{(mas)}  & 
\colhead{}  & 
\colhead{}  & 
}
\startdata
0.55 -- 0.85   & $2.56 \pm 0.04$   & \cite{Baines2023} \\
0.65 -- 0.85   & $2.47 \pm 0.05$   &  \cite{Nordgren1999}\\
1.45 -- 1.8   & $2.08\pm 0.01$    &  This work\\
1.60 - 2.20   & $2.15 \pm 0.04$   &  \cite{vanBelle2009}\\
1.98 -- 2.30  & $1.74\pm0.08$     &  \cite{Touhami2012} \\
1.98 -- 2.30  & $2.09\pm0.02$     &  This work \\
2.3 -- 2.38   & $2.91\pm 0.19$    &  This work \\
\enddata
\end{deluxetable*}

\begin{figure}[h]
\centering
\includegraphics[width=0.85\textwidth]{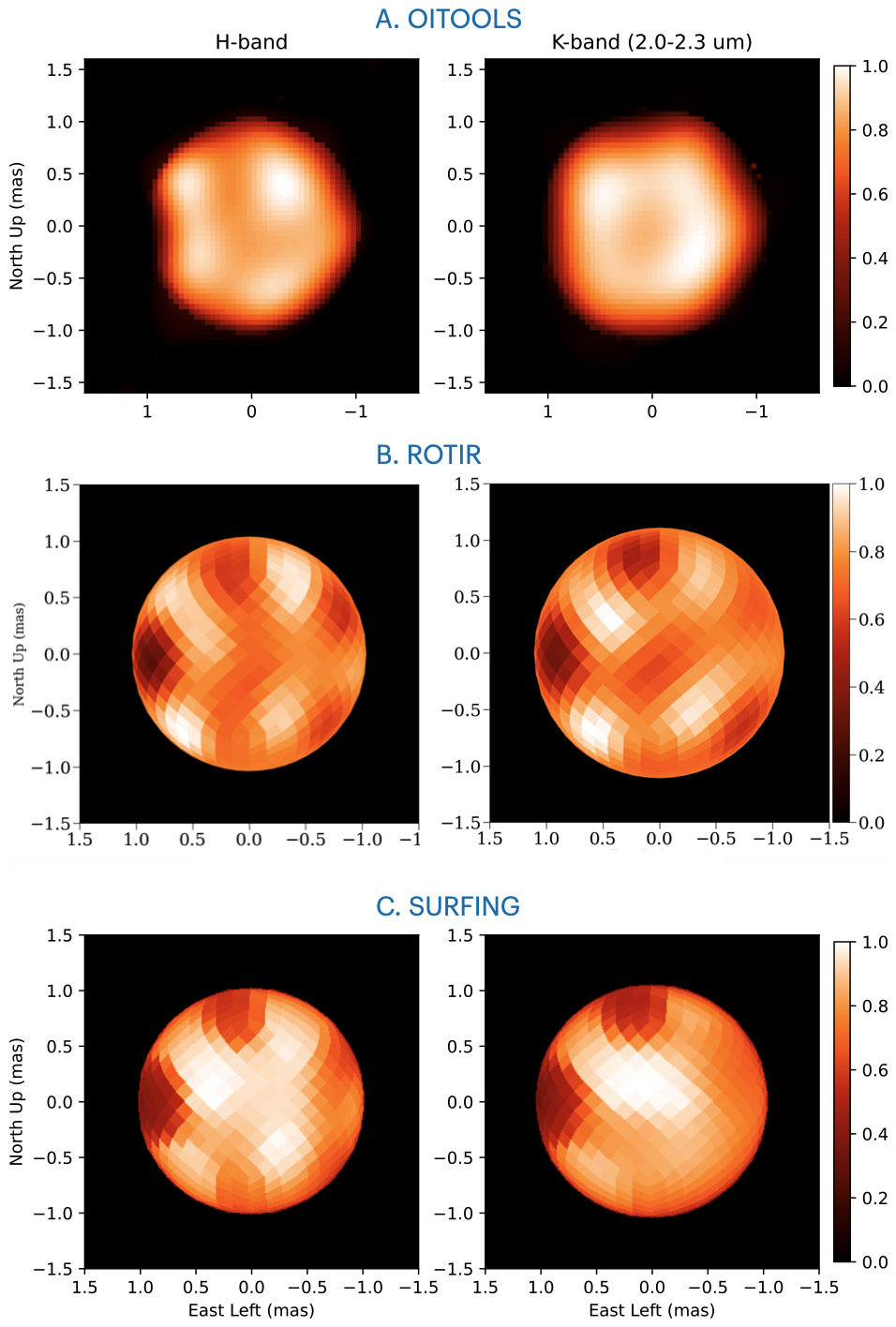}
\caption{These images were generated using the software OITOOLS (top), ROTIR (middle) and SURFING (bottom).  The left and right panels show H-band and K-band images, respectively. The image reconstructions are limited to the continuum. The intensity values in the images of each software normalized to maximum and are plotted linearly.  
}
\label{Fig:Final_images_merged}
\end{figure}

\begin{figure}[h]
\centering
\includegraphics[width=\textwidth]{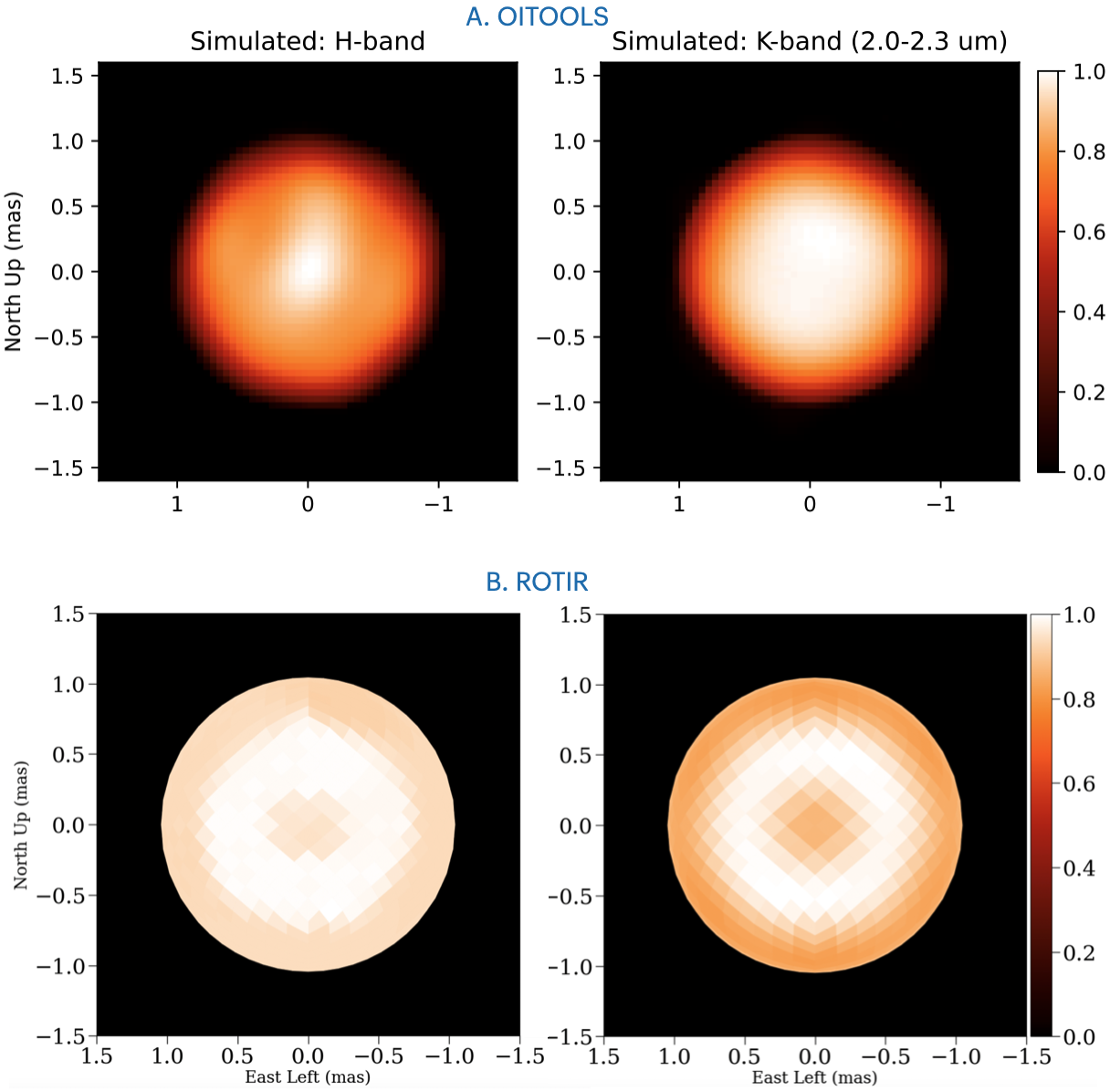}
\caption{To quantify the fidelity of the reconstructions shown in Figure~\ref{Fig:Final_images_merged}, analogous image reconstructions are carried out using synthetic observations. These synthetic observations mimic real observations of a limb-darkened star with a diameter of $\theta=2.1$ mas and replicate the same $(u,v)$ coverage. 
The top and bottom panels show the synthetic image reconstructions of OITOOLS and ROTIR. 
The intensities or the colors of the images are scaled to those in Figure~\ref{Fig:Final_images_merged}. Any features that depart from the general morphology shown here probably represent the actual spatial intensity distribution.
}
\label{Fig:Synthetic_images}
\end{figure}

\section{Image Reconstruction in Continuum Regions} \label{Sec:images}

In the second phase of our data analysis, we concentrated on creating model-independent aperture synthesis images of the continuum wavelength spectrum data. The incomplete $(u,v)$ coverage from the interferometry presents an ill-posed inverse problem, hindering the ability to derive a unique image from the measured set of visibilities. The reconstruction of images involves the minimization of a cost function comprising both data and regularization terms. The data consist of squared visibility $V^2$ and closure phase  $\phi$ and the regularization term incorporates a set of a priori assumptions about the target, such as positivity and smoothness.  We merged all MIRC-X and MYSTIC epochs into respective combined datasets to perform image reconstructions on the consolidated  $(u,v)$-coverage data sets. Appendix~\ref{Sec:epoch_images} shows the image reconstructions from the best ($u,v$)-coverage individual epochs. 

\newpage
\subsection{Strategy of the Image reconstructions}

We employed a diverse range of imaging techniques for these image reconstructions for the purpose of enhancing the reliability of the reconstructed images: 
(i) OITOOLS \citep{Norris2021,Martinez2021}\footnote{\href{https://github.com/fabienbaron/OITOOLS.jl.git}{https://github.com/fabienbaron/OITOOLS.jl.git}}, 
 (ii) ROTational Image Reconstruction (ROTIR) \citep{Martinez2021}\footnote{\href{https://github.com/fabienbaron/ROTIR.jl.git}{https://github.com/fabienbaron/ROTIR.jl.git}} and,  
 (iii) SURFace imagING (SURFING) \citep{Roettenbacher2016,Roettenbacher2017,Martinez2021,Anugu2023}. OITOOLS and ROTIR are written in Julia, while SURFING is written in the IDL programming language. 

OITOOLS utilizes a quasi-Newton method that incorporates rapid gradient minimization through Variable Metric Limited Memory with Bounds (VMLMB). This image tool is specifically designed for two-dimensional imaging applications.

On the other hand, ROTIR and SURFING are specialized in imaging three-dimensional surfaces of stars.  These techniques assign specific intensity values to each element on the three-dimensional surface of a star, allowing for the creation of detailed stellar surfaces. Parameters such as angular diameter, equatorial to polar axis ratio, inclination, position angle of the rotation pole, limb-darkening coefficient, rotation period, epoch, and visibility at a spatial frequency of 0 are considered in the fitting.

The initial step in image reconstruction involved determining the best-fit power-law limb-darkened angular diameter, which sets the outer boundary for the assigned flux. The best-fit diameters, denoted $\theta = \theta_{\rm LDD}$, are presented in Table~\ref{Table:LDD_fit}. 

In OITOOLS, we applied positivity constraints and edge-preserving regularization $\ell_2-\ell_1$. The regularizer weight (hyperparameter) was optimized using the $L$-curve method. To mitigate the effects of local minima, we initiated 1000 reconstructions from random initial generated images. We also searched a range of pixel scales (0.05 mas to 0.2 mas) and fields of view (6.4 mas to 12 mas) and  eventually settled on pixel scale 0.05 mas,  with a final field of view of $6.4\times6.4$~mas. Despite varying different regularizers, a consistent large-scale asymmetry was observed in the images (top row in Figure \ref{Fig:Final_images_merged}).  For additional verification, we reconstructed images from data averaged over 10-minute integrations, assessing their robustness against those from shorter intervals (150 s). Both integration sets gave the nearly the same images.

In ROTIR and SURFING, the star is assumed to be a sphere, so the ratio of the equatorial to polar axes is set to 1. 
The rotational distortion of the star is probably small.  We computed the Roche figure for the mass, polar radius, and projected equatorial velocity given in Table 1, and we found that the difference between the projected major and minor axes is less than $3\%$. 
We fix the stellar inclination and the position angle of the axis of rotation in the global model because the time between the epochs of observation is significantly less than the approximately 3-year rotation period expected for $\rho$~Cas.  We arbitrarily set the epoch to the first night of our observation.

For ROTIR spheroid image reconstruction, we fitted several stellar parameters with initial uniform prior distributions within a wide range of values taken from Table~\ref{Table:LDD_fit}: the angular size of the radius at the pole (1.05 mas), surface temperature, and limb-darkening power law coefficient ($\alpha_{\rm H}\sim0.47$ and $\alpha_{\rm K}\sim0.38$, Figure~\ref{Fig:oitools_LDD_fit}). We used a total variation regularizer with weight of 0.05. Figure~\ref{Fig:Final_images_merged} middle row, shows the ROTIR images.

The SURFING software fits the stellar diameter as part of the image reconstruction process, which resulted in diameters of 2.04 mas and 2.10 mas for the H-band and K-band wavelengths, respectively. We used fixed limb-darkening coefficients ($\alpha_{\rm H}\sim0.47$ and $\alpha_{\rm K}\sim0.38$) derived from Section~\ref{Sec:ldd_fit}.  Figure~\ref{Fig:Final_images_merged} bottom row, shows the SURFING images.

Figure~\ref{Fig:Final_images_merged} displays the images from the OITOOLS (top row), ROTIR (middle row) and SURFING (bottom row) software, respectively. A visual inspection of the images reveals that there are common characteristics and some differences between them. We were concerned whether the ring-shaped bright spots observed in the images (Figure \ref{Fig:Final_images_merged}) are real or result from the incomplete $(u,v)$ coverage. To explore how the reconstructed images were affected by various imaging and optimization choices, we generated synthetic data sets. The synthetic OIFITS datasets were generated numerically using a limb-darkened disk model based on the power law coefficient and with the angular diameter $\theta_{\rm LDD}=2.1$ mas measured from the observed data (from Table~\ref{Table:LDD_fit}) and also mimicking our actual observations, including the coverage $(u,v)$ and the signal to noise ratios. The image reconstructions were generated from these datasets and these did not show same image patterns (Figure~\ref{Fig:Synthetic_images}), indicating that the light and dark patterns are not an artifact of the incomplete coverage $(u,v)$. 
Appendix~\ref{Sec:epoch_images} displays individual epoch image reconstructions that exhibit excellent agreement with the image reconstructions derived from the combined data.

\begin{figure}[h!]
\centering
\includegraphics[width=0.49\textwidth]{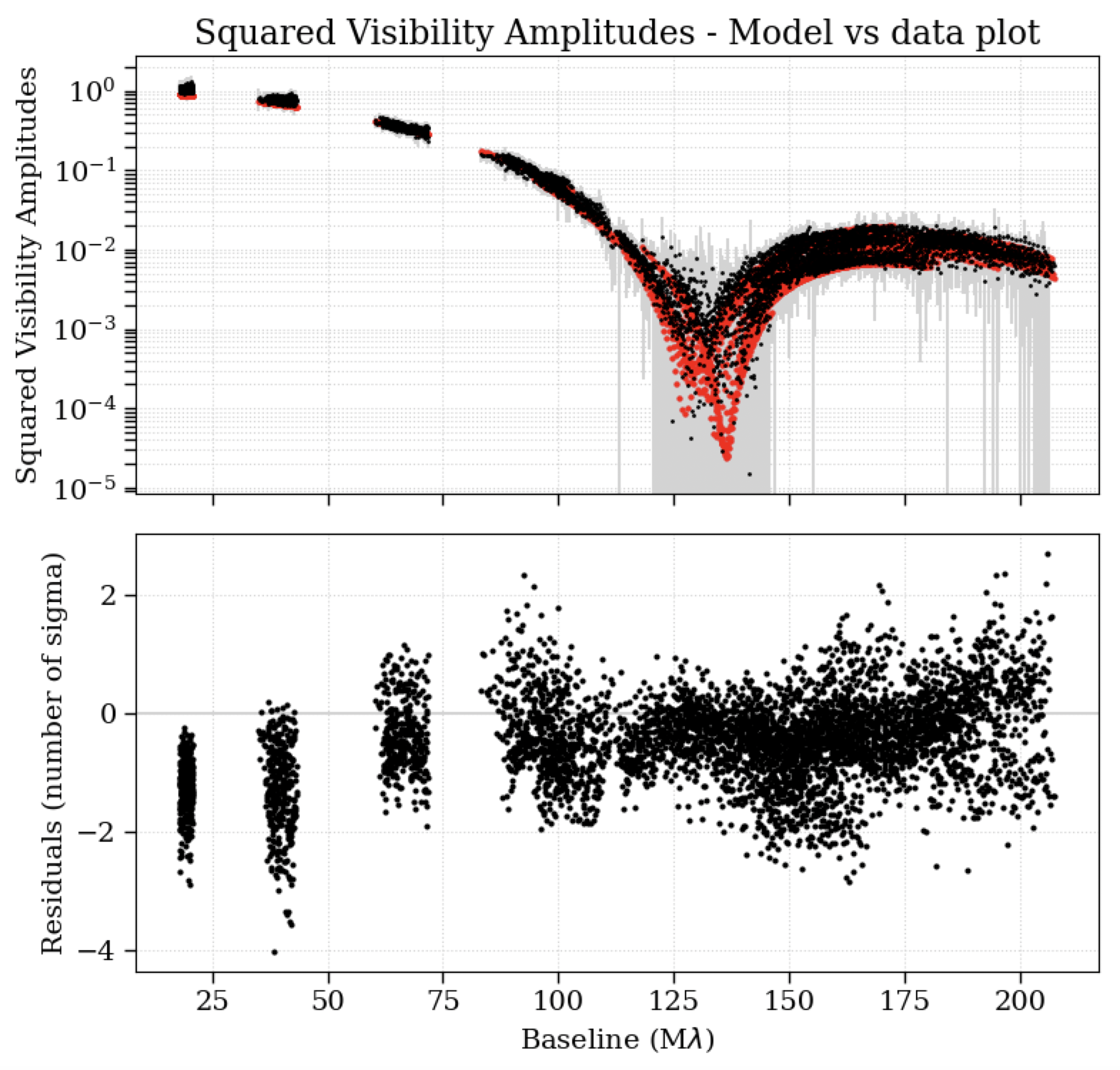}
\includegraphics[width=0.49\textwidth]{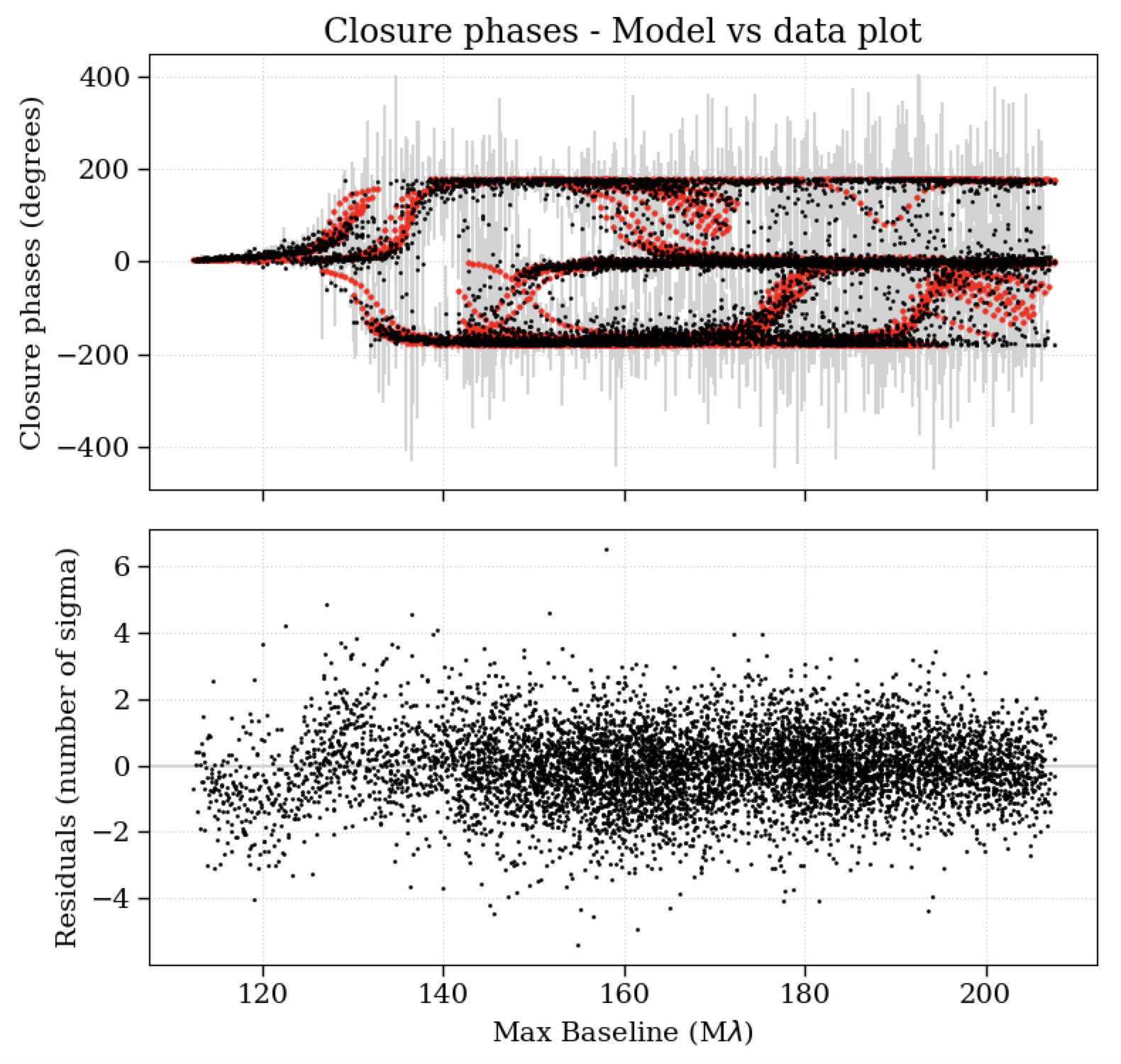} 
\caption{  
The ROTIR fit residuals to the MIRC-X (H-band) interferometric observables. In the top panels, the black and gray points are the observed data and uncertainties, respectively. The red colored points are the model fit. The bottom panels are the best-fit residuals of $V^2$ and $\phi$.  
}
\label{Fig:residuals_mircx}
\end{figure}

\begin{figure}[h!]
\centering
\includegraphics[width=0.49\textwidth]{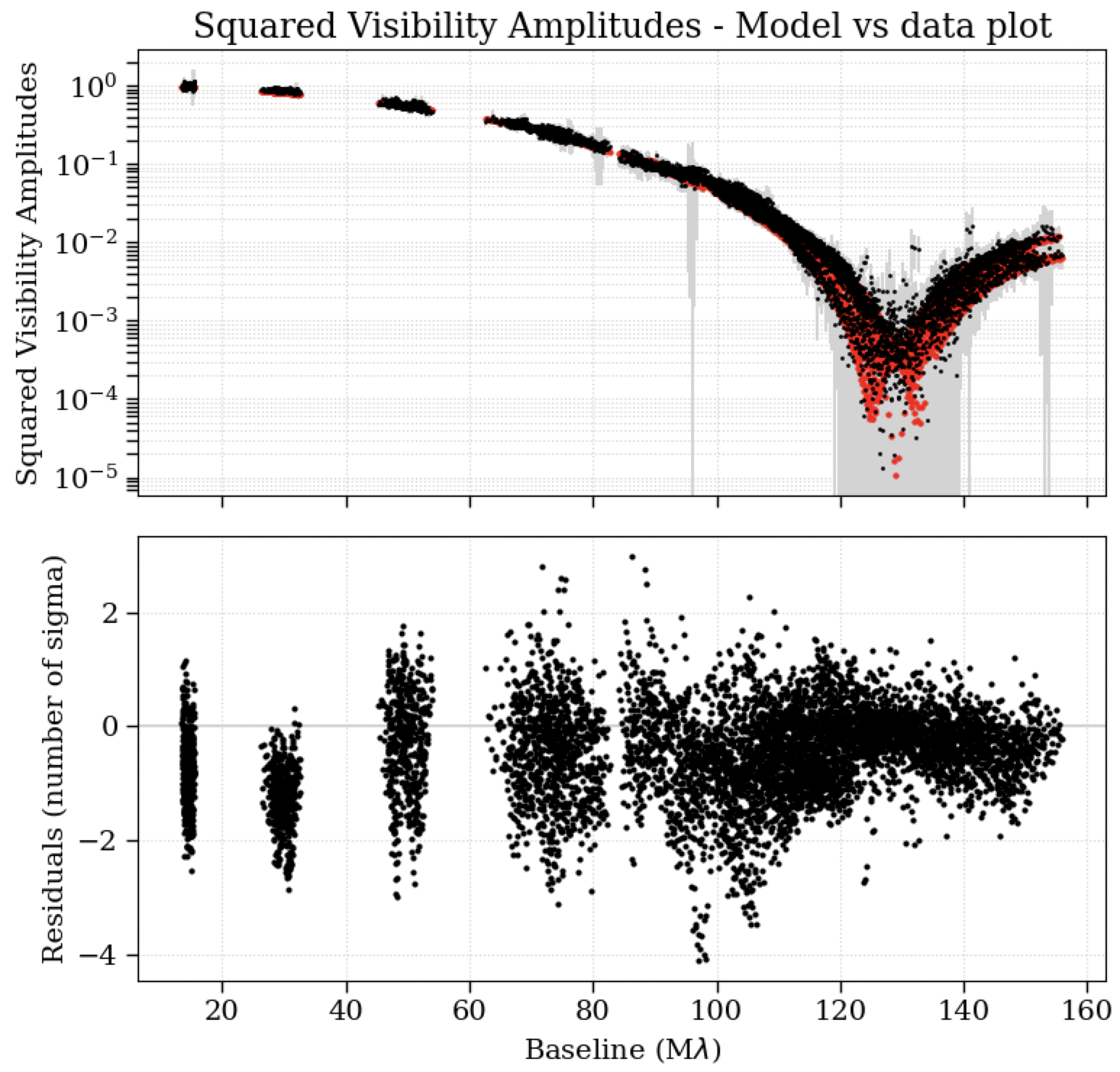}
\includegraphics[width=0.49\textwidth]{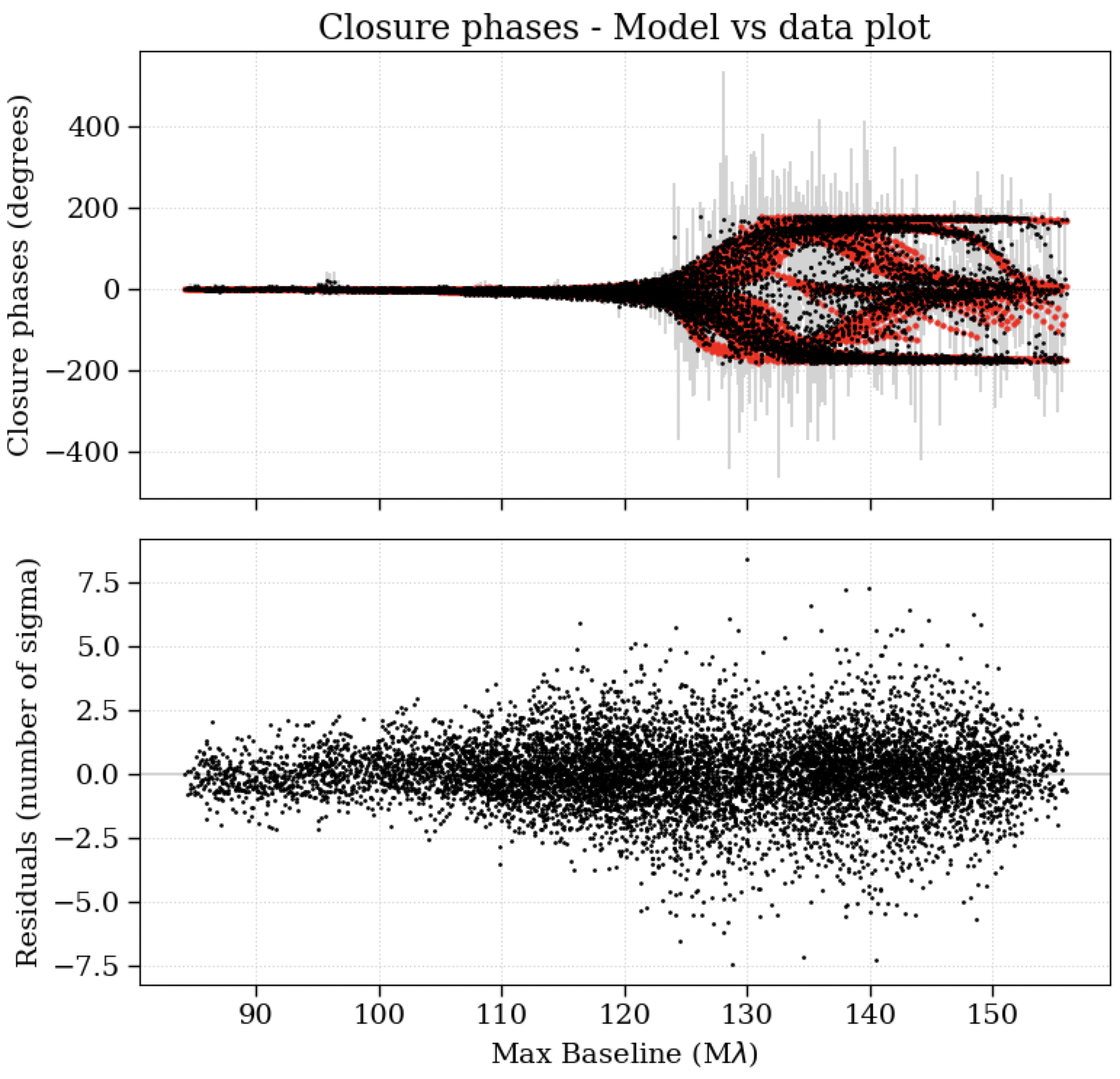}
\caption{  
The ROTIR fit residuals to the MYSTIC (K-band) interferometric observables. Colors are same as in Figure~\ref{Fig:residuals_mircx}.
}
\label{Fig:residuals_mystic}
\end{figure}

Figures~\ref{Fig:residuals_mircx} and \ref{Fig:residuals_mystic} show the fitting residuals for the ROTIR image reconstruction. The fitting residuals for  ROTIR, OITOOLS and SURFING are also qualitatively similar in appearance. The reduced $\chi^2$ values were 1.3 for MIRC-X data fits and 1.2 for MYSTIC data fits. We did not perform image reconstructions with the full K-band wavelengths, including the CO lines. However, we isolated the part of the spectrum recording the CO lines ($\lambda=2.3-2.38~\mu$m) and conducted image reconstructions that we discuss in Section~\ref{Sec:CO-lines}.

\subsection{Comparison between imaging results}

A visual inspection of the images (Figure~\ref{Fig:Final_images_merged}) obtained using the  OITOOLS,  ROTIR  and SURFING software shows both similarities and significant differences among them. The similarities between the images generated by OITOOLS, ROTIR and SURFING are particularly noteworthy, especially in terms of the positions of the bright and dark spots. This consistency across different image reconstruction techniques reinforces the reliability of the results and suggests that the bright and dark spots present in the images are significant and persistent features on the star's surface.

The bright spots in the OITOOLS image exhibit a ring-like shape, which differs somewhat from those obtained using ROTIR and SURFING. This dissimilarity could be attributed to the distinct underlying assumptions these algorithms make regarding the expected appearances of the stellar surface. OITOOLS operates with fewer assumptions about the star's appearance. In contrast, the ROTIR and SURFING images are based on the assumption that the star has a spheroid shape.  

Additionally, we noted apparent differences between the images captured in the H and K-band wavelength bands. These differences can be largely attributed to the variations in the effective angular resolutions of the telescope for each band. The H-band observations were more resolved than those in the K-band, leading to images with greater contrast and detail across the stellar disk.

\begin{figure}[h]
\centering
\includegraphics[width=\textwidth]{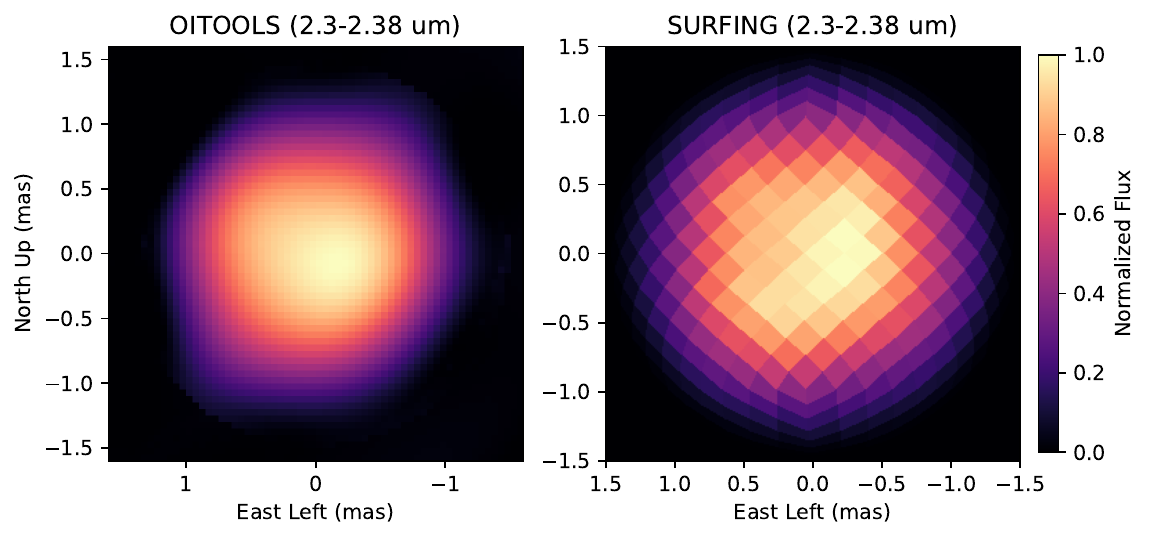}
\caption{Image reconstructions of the spectral region containing the CO spectral lines ($\lambda=2.3-2.38~\mu$m) are shown in the left and right panels from OITOOLS and SURFING, with intensities plotted on a linear scale. The central bright region is the photosphere, and outer region comprises the circumstellar envelope that appears larger because of the optical depth of the CO lines.
}
\label{Fig:CO_lines}
\end{figure}

\section{Image Reconstruction in the CO spectral lines} \label{Sec:CO-lines}

We isolated the CO lines ($\lambda=2.3-2.38~\mu$m) from the K-band data and performed image reconstructions using OITOOLS and SURFING, employing the same methods outlined in the previous section. Figure~\ref{Fig:CO_lines} shows the images, revealing two components: a central star photosphere and an outer circumstellar envelope. The OITOOLS image reveals an asymmetric shape of the circumstellar envelope that appears smaller along the North-East to South-West direction. The OITOOLS and SURFING software measure the outer diameter of the circumstellar envelope as $2.91 \pm 0.19$ mas and $2.96$ mas, respectively.  We have higher confidence in the OITOOLS image in this case compared to the SURFING software, because the circumstellar envelope does not necessarily need to be a spherical shape. ROTIR was not implemented for these particular data, as it also enforces a spherical shape, which may not be suitable for this asymmetric shape of the circumstellar envelope.


\section{Discussion} \label{Sec:discussion}

\subsection{Physical size of photosphere and circumstellar envelope}

Our LDD model fitting reveals that the angular diameter of the star $\rho$~Cas is approximately $\theta_{\rm LDD} \sim 2.1$ mas, as shown in Table~\ref{Table:LDD_fit}. Converting this angular size into physical units requires an accurate distance measurement, but the exact distance to $\rho$~Cas remains under debate. \cite{Zsoldos1991A&A...246..441Z} suggested a distance of about $3.1\pm0.5$ kpc, based on interstellar absorption and absolute visual magnitude calculations. \cite{Kusuno2013ApJ...774..107K} estimate a distance of $2810^{+220}_{-190}$~pc based on membership in the Cas OB5 association. \cite{vanGenderen2019} estimate $2.5\pm0.3$ kpc based on the stellar radii monitoring from the spectroscopic and radial velocity observations during the eruption in the year 2000.  These estimates are inconsistent with measurements from the Gaia mission. Gaia DR2 data suggests a closer distance of approximately $1.1^{+0.4}_{-0.2}$ kpc \citep{BailerJones2018}, while more recent Gaia EDR3 data indicates a much greater distance of about $6.7^{+1.9}_{-1.6}$ kpc \citep{BailerJones2021}. The significant discrepancies in Gaia measurements could be attributed to biases in astrometric measurements related to the star's large brightness and potential photocenter jitter caused by the stellar convection cells \citep{Chiavassa2022}. In this study, we disregard the Gaia measurements because of their inconsistency. We consider a distance range of 2.5 to 3.1 kpc for $\rho$~Cas. Using the photosphere angular diameter $\theta_{\rm LDD} = 2.1$ mas, extracted from OITOOLS images, we calculated the radius of $\rho$~Cas to be in the range of $564 \pm 67$ to $700\pm112~R_\odot$ or $2.6\pm0.3$ to $3.3 \pm 0.5$ au. This measured value slightly exceeds the earlier estimate of $450 ~R_\odot$ derived from considerations involving luminosity and effective temperatures \citep{Stothers2012}. 

Table~\ref{Table:LDD_fit} presents a comparison of historically measured diameters of $\rho$~Cas. Our measured diameter ($2.1 \pm 0.02$ mas) matches  with \cite{vanBelle2009} within measurement errors. However, there is a significant discrepancy between our result and the measurements obtained by \cite{Touhami2012} during the 2006-2009 period. Their reported diameter is much smaller ($1.74\pm0.08$ mas), which might be attributed to a presence of a central, brighter spot during their observations, suggesting potential changes in the convection cell structure over time.

Table~\ref{Table:rhoCas_YHGs_compare} compares the fundamental parameters of $\rho$ Cas with other established YHGs. When looking at size based on interferometric observations, $\rho$ Cas falls in the middle range. HR~5171~A is larger than $\rho$ Cas, while both IRC+10420 and HR~8752 are smaller.

\begin{deluxetable*}{ccccccccc}[h]
\tablecaption{Comparison of size of $\rho$ Cas with the established YHGs \label{Table:rhoCas_YHGs_compare} }
\tablewidth{0pt}
\tablehead{
\colhead{Object} & 
\colhead{Wavelength} & 
\colhead{Ang diameter} & 
\colhead{Distance} & 
\colhead{Radius} & 
\colhead{Reference} 
\\ 
\colhead{Name}  & 
\colhead{($\mu$m)}  & 
\colhead{(mas)}  & 
\colhead{(kpc)}  & 
\colhead{($R_\odot$)}  & 
\colhead{}  
}
\startdata
$\rho$ Cas & 1.45 -- 2.30  & $2.09\pm0.02$ &   2.5-3.1      & $564\pm67$ to $700\pm112$ & This work \\
HR~5171~A & 1.45 -- 1.80  & $3.39 \pm 0.02$ &   $3.6\pm0.5$      & $1315 \pm 260$ & \cite{Wittkowski2017} \\
HR~8752 & 1.45 -- 1.8  & $1.22 \pm 0.03$ &   $3.9\pm 0.8$      & $511 \pm 112$ & \cite{vanBelle2009} \\
IRC+10420 & 1.9 -- 2.30  & $0.70-0.98$ &   $4.2\pm 0.8$     & $442 \pm 84$ & \cite{Oudmaijer2013} \\
RW~Cep   &1.45 -- 2.30 & $2.45\pm 0.04$  &   $3.4 - 6.7$ & 900 - 1760 & \cite[][Anugu et al. prep]{Anugu2023}\\
\enddata
\end{deluxetable*}

\subsection{Convection cells and mass loss}\label{Sec:mass_loss}

The largest bright spot in our reconstructed images (Figure~\ref{Fig:Final_images_merged}), which makes up about 25\% of the star's size, indicates a radius of $141 - 175~R_\odot$. We propose that the four or five bright spots observed in the reconstructed images likely represent giant convection cells, while the two dark regions in the images (North and East) correspond to the cool infalling material situated between these giant convection cells. \cite{Norris2021} monitored the red supergiant AZ Cyg for five years with the CHARA Array in H-band with the MIRC instrument and found that AZ Cyg exhibits a configuration of large and long-lived convection cells lasting more than a year, alongside smaller-sized hot granules that endure for months but last less than a year.  In particular, HR~5171~A is the only other yellow hypergiant with interferometric images of the stellar surface and circumstellar envelope in the near-infrared  \citep[in H-band,][]{Chesneau2014A&A...563A..71C, Wittkowski2017}.
\citet{Wittkowski2017} made three epochs of observations spanning three years, and they imaged the surface and discovered an eclipsing companion in contact with the primary yellow hypergiant. Although two convection cells were observed in the first epoch (2014), only one was present in the later epochs (2016 and 2017), separated by two years. The star $\rho$~Cas, with high angular resolutions offered by CHARA, provides another target of opportunity for a long-term monitoring study to conduct a comprehensive life cycle analysis of these convection cells.

Historically, $\rho$~Cas has experienced significant episodes of mass loss, leading to noticeable variations in visual brightness. These fluctuations, marked by dimming and re-brightening patterns \citep{Beardsley1961ApJS....5..381B, Leiker1989IBVS.3345....1L, Lobel2003ApJ...583..923L, Kraus2019MNRAS.483.3792K}, are similar to those observed in the Great Dimming of Betelgeuse \citep{Montarges2021, Dupree2022} and RW Cephei \citep{Anugu2023}. Our long-wavelength K-band observations ($\lambda > 2.3 \mu$m) have unveiled evidence of extended structures, as illustrated in Figure~\ref{Fig:CO_lines}. These structures, approximately $2.91\pm0.19$ mas in size, suggest the presence of molecular emission lines emanating from CO shells surrounding the star's photosphere.  Historically, $\rho$~Cas has displayed both emission and absorption spectral lines, exhibiting a correlation with its visual light curve.  \cite{Gorlova2006} and \cite{Yamamuro2007PASJ...59..973Y} compared the CO-line spectra with the visual light curve, revealing a pattern of absorption, non-detection, and emission during the mass loss events. The CO emission lines consistently appeared during phases of maximum light, returning to CO absorption lines shortly after the star dimmed or experienced an outburst. During our CHARA observations, there was no visual dimming (nor outburst), and we measured the presence of emission lines in our photometric datasets (See Appendix~\ref{Sec:OIFLUX}).

From the recent studies of the dimming of Betelgeuse \citep{Montarges2021, Dupree2022} and RW Cep \citep{Jones2023,Anugu2023}, it is evident that episodic mass loss events in massive evolved stars are linked to the activity of large and hot giant convection cells on the stellar surface.  The convection cells found in the reconstructed images may have played a role in the episodic mass loss events recorded for $\rho$~Cas previously. Further substantiation of this connection could be achieved through observations revealing the dynamical evolution of the convection cells through time-resolved observations or/and Doppler velocity measurements of the surface.  While potential differences in dust chemistry and grain size between red supergiants and yellow hypergiants could aid our understanding of the dimming phenomenon, a detailed analysis of YHG dust properties is beyond the scope of this current work. However, to provide some insights, we compare the dust grain sizes measured for Betelgeuse and IRC+10420 (the closest yellow hypergiant analogue to $\rho$~Cas). Interestingly, the silicate dust grain sizes appear to be quite similar. Betelgeuse hosts grains ranging from $0.05 - 0.5\mu$m \citep{Mauron1986A&A...168..217M}, with an average size of $\sim0.3 \mu$m \citep{Haubois2019A&A...628A.101H}. Similarly, IRC+10420 exhibits dust grains within the range of $0.005 - 0.45 \mu$m \citep{Blocker1999A&A...348..805B}.

\cite{Humphreys2022} argued that red supergiants rely not only on convection but also on the influence of magnetic fields, which play a vital role in driving the surface mass ejections. These magnetic fields propel gas clouds into space and contribute to the creation of dust and circumstellar envelopes. However, in the specific case of $\rho$~Cas, \cite{Tessore2017A&A...603A.129T} was not able to validate the presence of magnetic fields, but placed only upper limits on the magnetic field strength.

\subsection{The occurrence of secondary long periods driven by giant convection cells?}

The discovery of these giant convection cells in $\rho$~Cas adds an intriguing dimension to the investigation of the mechanisms driving the variability in yellow hypergiants. Among these stars, two distinct types of variability are observed: shorter primary periods  and longer secondary periods  \citep{Stothers1971A&A....10..290S,Kiss2006MNRAS.372.1721K}. The shorter primary periods are commonly attributed to radial pulsation \citep{Stothers1969ApJ...156..541S}, while the long secondary periods are believed to be linked to the turnover time of giant convection cells in a stellar envelope characterized by significant depth and strong convection \citep{Stothers2012}. The detection of giant convection cells in $\rho$~Cas makes it an ideal target to study the connection between the convection cells and its secondary long periods. $\rho$~Cas displays shorter periods with irregular range from 200 - 500 days and a secondary period of 820 days \citep{Percy2000}.   Conducting further observations covering the 820-day period could serve to test and validate a causal connection between the convection cells and the long period variability and mass loss (see Section~\ref{Sec:mass_loss}). 

\subsection{Limb brightening?}

Convection cell activity can help explain the appearance of the star in
the reconstructed images, but the appearance of the bright zones towards
the stellar limb in all the reconstructions is puzzling (Appendix~\ref{Sec:limb_darkening_analysis}).  Alternatively, 
the brighter outer regions may be the result of heating and hotter temperatures
in the star's outer atmosphere.  The classical limb darkening observed in
the Sun and other stars is interpreted through the Eddington-Barbier relationship
between the specific intensity $I(\mu)$ and the source function $S(\tau)$.
The drop in intensity from center to limb then reflects the decrease in
temperature and smaller source function higher in the atmosphere (at lower
optical depth $\tau$).  Instead, the apparent relatively brighter limb found in
the images of $\rho$~Cas would indicate that the atmospheric temperature rises
at some level higher in the photosphere.  We suggest that such heating could be
the result of pulsation-induced shocks.  The star is a well-known semi-regular variable with pulsational cycles in the range of 200 to 500 days \citep{vanGenderen2019}. The associated radial velocity variations are larger than the sound speed \citep{Lobel2003ApJ...583..923L}, and this suggests that the pulsations could create shocks
and shock heating in the upper atmosphere \citep{Wood1979ApJ...227..220W,Bowen1992iesh.conf..104B}.

\cite{deJager1997A&A...325..714D} developed a model for shock propagation
in the atmosphere of $\rho$~Cas that helps explain the large ``microturbulent''
broadening of spectral lines and the variable H$\alpha$ emission. 
Specifically, they show (in their Fig.\ 4) how the shocks increase the gas
temperature in the upper atmosphere above optical depth unity.  
While they do not discuss the implications of this heating for the variation in
continuum intensity across the face of the star, a trend of rising temperature
at smaller optical depths is consistent with a brightening towards the limb
as observed.  \cite{deJager1998A&ARv...8..145D} suggests that one shock zone (associated with one
pulsational cycle) will dominate the heating in the atmosphere at any given time.
Thus, the amplitude and height of the shock heating may vary throughout the
pulsation cycle, and future interferometric observations may offer a test of
this model by measuring the changes in limb brightening throughout the cycle.

\section{Summary}

In this paper, we present high-angular-resolution ($\sim0.5$ mas) observations of the yellow hypergiant $\rho$~Cas that we made using the CHARA Array, employing the MIRC-X and MYSTIC instruments in H- and K-band wavelengths. Fundamental parameters of $\rho$~Cas were determined using two model-fitting techniques and three different image reconstruction techniques analyzing the data. Our model fitting  constrains the stellar diameter to $2.09\pm 0.02$~mas. Assuming a distance range of 2.5-3.1 kpc, $\rho$~Cas has a radius of $564-700~R_\odot$ in the near-infrared. 

The primary result of our study is the first surface imaging of $\rho$~Cas, showcasing unprecedented details of its photosphere. These images reveal four to five  bright spots, which we interpret as manifestations of convection cells. We suggest that these convection cells might have played a pivotal role in ejecting gas clouds from the stellar surface, thereby contributing to the episodic mass loss events recorded previously. 

Our analysis also shows that the $V^2$ and $\phi$ data appear different in the CO bandhead emission lines ($\lambda=2.31-2.38 ~\mu$m), leading to the detection of an extended circumstellar envelope of size with a diameter of $2.91\pm0.19$ mas. Such envelopes are characteristic of cool and luminous stars. 

Future follow-up observations of $\rho$~Cas with the CHARA Array will provide a comprehensive monitoring of the dynamics of the convection cells on the star's photosphere. Furthermore, linking these observational data with future episodes of stellar dimming will be crucial to advance our understanding of the mass loss phenomena and their driving mechanism in massive stars. 
We are planning to continue the CHARA Array observations to explore the correlation between the evolving convection cells in the images and the associated photometric variations.

Yellow hypergiants are extremely rare; currently, only about 10 objects have been identified in our Milky Way Galaxy. In particular, only five of these objects ($\rho$~Cas, HR~8752, IRC~+10420, HD~179821, RW~Cep) are accessible for observation in the northern sky. Among them, $\rho$ Cas and RW~Cep are only large enough for stellar surface imaging, making it crucial for follow-up imaging studies. We aim to extend CHARA observations to the remaining yellow hypergiants to measure their physical diameters and gain a detailed understanding of these objects.

\vspace{0.5 cm}
{\nolinenumbers
We express our sincere gratitude to the anonymous reviewer for their valuable and constructive comments, which contributed to the improvement of the paper. We also thank Stephen Ridgway for his insightful feedback, which strengthened the manuscript.
This work is based upon observations obtained with the Georgia State University
Center for High Angular Resolution Astronomy Array at Mount Wilson Observatory.  
The CHARA Array is supported by the National Science Foundation under Grant No.\
AST-1636624, AST-1908026, and AST-2034336.  Institutional support has been provided 
from the GSU College of Arts and Sciences and the GSU Office of the Vice President 
for Research and Economic Development. 
F.B.\ acknowledges funding from the National Science Foundation under Grant No.\ AST-1814777.
J.D.M.\ acknowledges funding for the development of MIRC-X (NASA-XRP NNX16AD43G, 
NSF AST-1909165) and MYSTIC (NSF ATI-1506540, NSF AST-1909165). 
R.M.R. acknowledges support from the Heising-Simons Foundation 51 Pegasi b Fellowship. 
S.K.\ acknowledges support from the European Research Council through a Starting Grant 
(Grant Agreement No.\ 639889) and Consolidator Grant (Grant Agreement ID 101003096).
M.M. acknowledges funding from the Programme Paris Region fellowship supported by the R\'egion Ile-de-France. This project has received funding from the European Union’s Horizon 2020 research and innovation program under the Marie Sk\l{}odowska-Curie Grant agreement No. 945298.
We acknowledge with thanks the variable star observations from the AAVSO International Database contributed by observers worldwide and used in this research.
This research has made use of the Jean-Marie Mariotti Center Aspro and SearchCal services.}

\facility{CHARA}
\software{PMOIRED \citep{antoine_merand_2024_10889235},  OITOOLS \citep{Norris2021}, ROTIR \citep{Martinez2021}, SURFING \citep{Roettenbacher2016}, mircx\_pipeline \citep{le_bouquin_2024_12735292}}

\appendix

\section{Image reconstructions for individual epochs}\label{Sec:epoch_images}

\begin{figure}[h!]
\centering
\includegraphics[width=\textwidth]{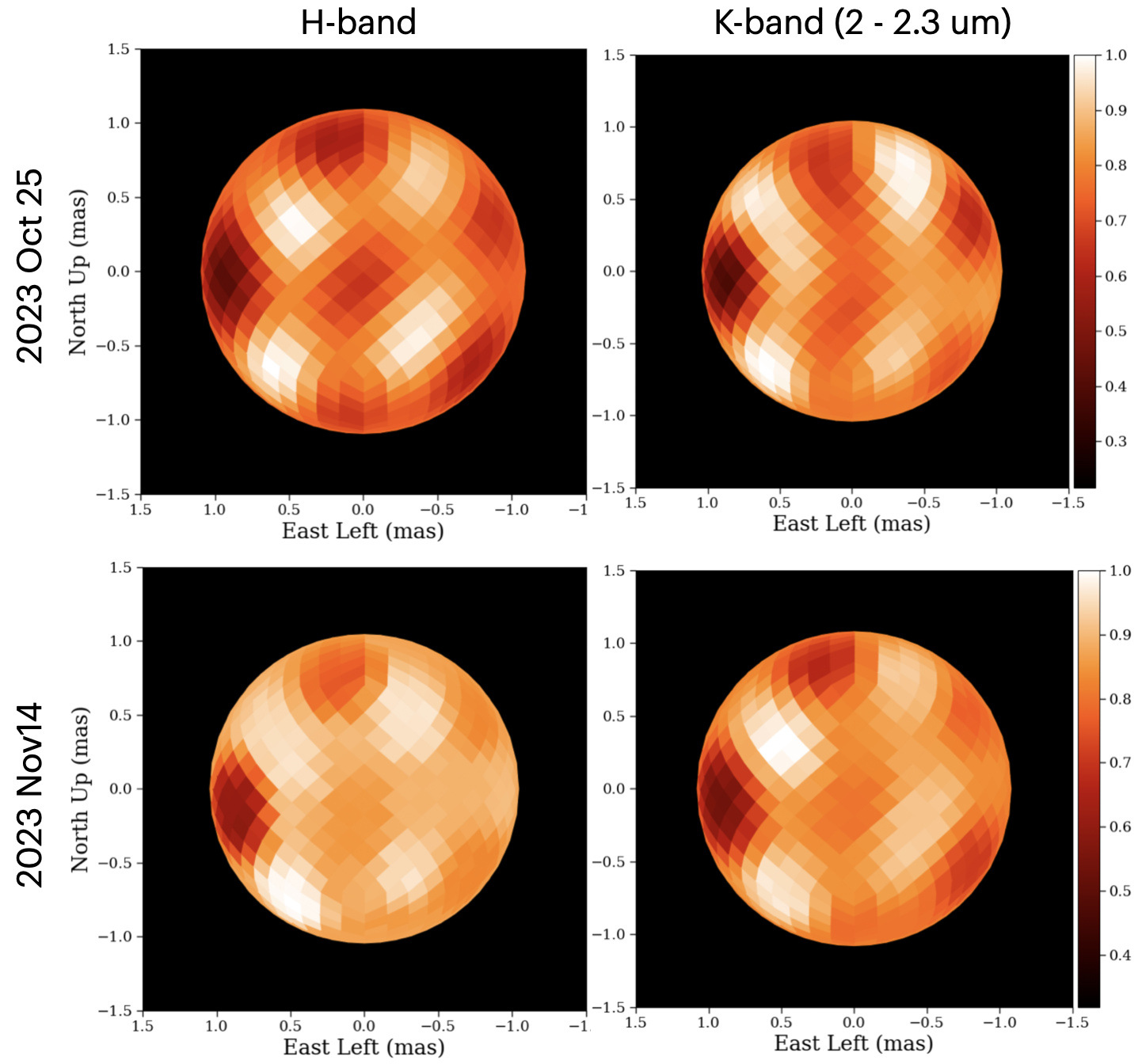}
\caption{Individual ROTIR image reconstructions to check the robustness of image reconstructions. The top and bottom panels are for the 2023 October 25 and 2023 November 14 epochs. The labels are same as in Figure~\ref{Fig:Final_images_merged}.}
\label{Fig:ROTIR_individual}
\end{figure}

\begin{figure}[h]
\centering
\includegraphics[width=\textwidth]{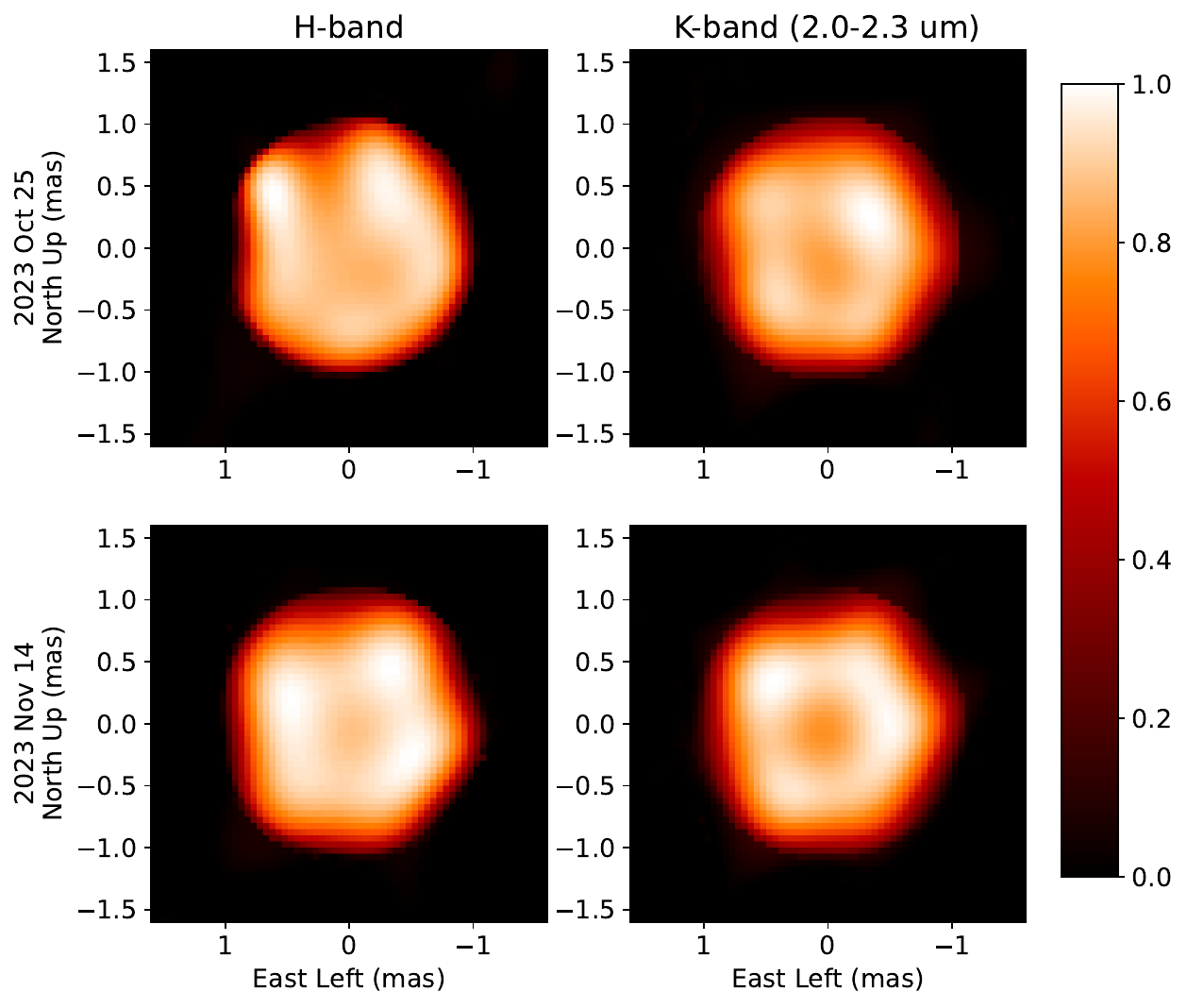}
\caption{Individual OITOOLS image reconstructions to check the robustness of image reconstructions. The top and bottom panels are for the 2023 October 25 and 2023 November 14 epochs. The intensities or the colors of the images are scaled to those of Figure~\ref{Fig:Final_images_merged}.}
\label{Fig:OITOOLS_individual}
\end{figure}

In order to evaluate the robustness of image reconstructions, we reconstructed individual epoch images for two nights that have good $(u,v)$-coverage data and higher signal-to-noise ratio: 25 October 2023 and 14 November 2023. The images from ROTIR and OITOOLS are depicted in Figures~\ref{Fig:ROTIR_individual} and \ref{Fig:OITOOLS_individual}. The individual epoch images exhibit good agreement with the combined images, reinforcing the reliability of the reconstruction process.

\section{Limb darkening from the image analysis}\label{Sec:limb_darkening_analysis}

\begin{figure}[h!]
\centering
\includegraphics[width=\textwidth]{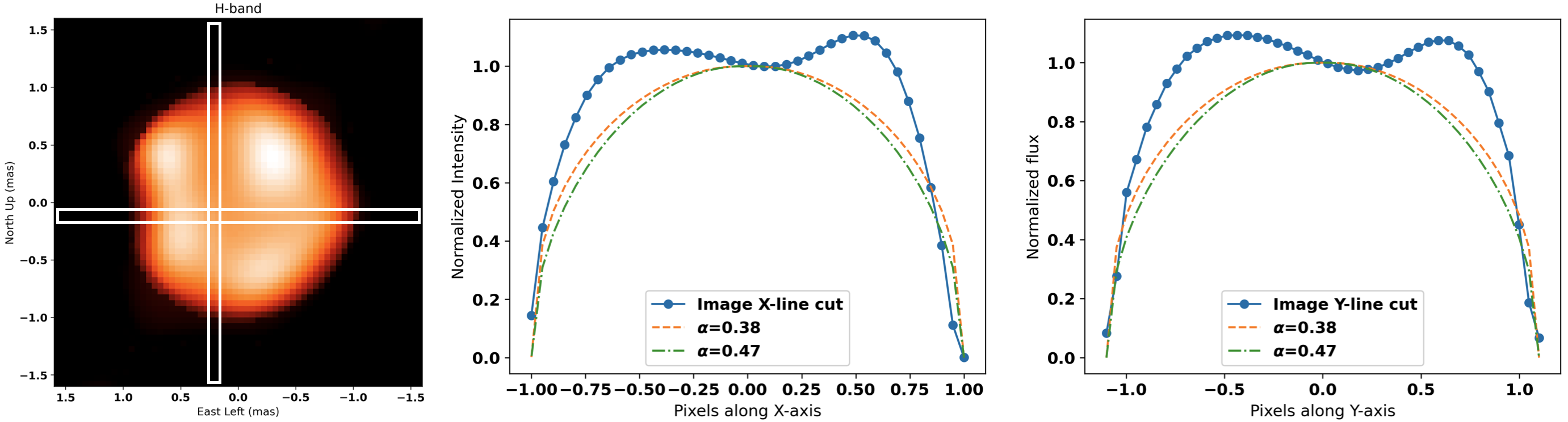}
\caption{  
In the left panel, we select 1-dimensional samples of pixels from the H-band image from OITOOLS (Figure~\ref{Fig:Final_images_merged}) for plotting. The middle panel presents the intensity of the star along the chosen horizontal strip line. The right panel presents the intensity of the star along the vertical strip line. The intensity at the center of the image is normalized to 1.0 to match the theoretical limb darkening for the intensity at the center of the star. In the middle and right panels, the filled circles represent the intensity data from the image. The dashed and dotted lines denote the measured limb darkening coefficients from Sections \ref{Sec:ldd_fit} and \ref{Sec:images}.
}
\label{Fig:ldd_flux}
\end{figure}

In Figure~\ref{Fig:ldd_flux}, the intensity values along the X and Y-axes are plotted based on the H-band OITOOLS image. We selected the region on the image where no bright spots arepresent. The intensity values at the center are initially low, gradually increase, and subsequently decrease towards the edges.  The intensity does not exhibit a smooth transition from the center to the edge, creating challenges in accurately fitting the limb darkening model to the observed data (Section \ref{Sec:ldd_fit}). 

In Figure~\ref{Fig:ldd_flux}, we over-plotted the limb darkening with power law coefficients $\alpha=0.47$ and $0.38$, measured in Section \ref{Sec:ldd_fit}, for comparison. Based on the differences between the observations and theoretical expectations, it may be that $\rho$~Cas is experiencing limb brightening.


\section{K-band flux analysis}\label{Sec:OIFLUX}
\begin{figure}[h!]
\centering
\includegraphics[width=0.5\textwidth]{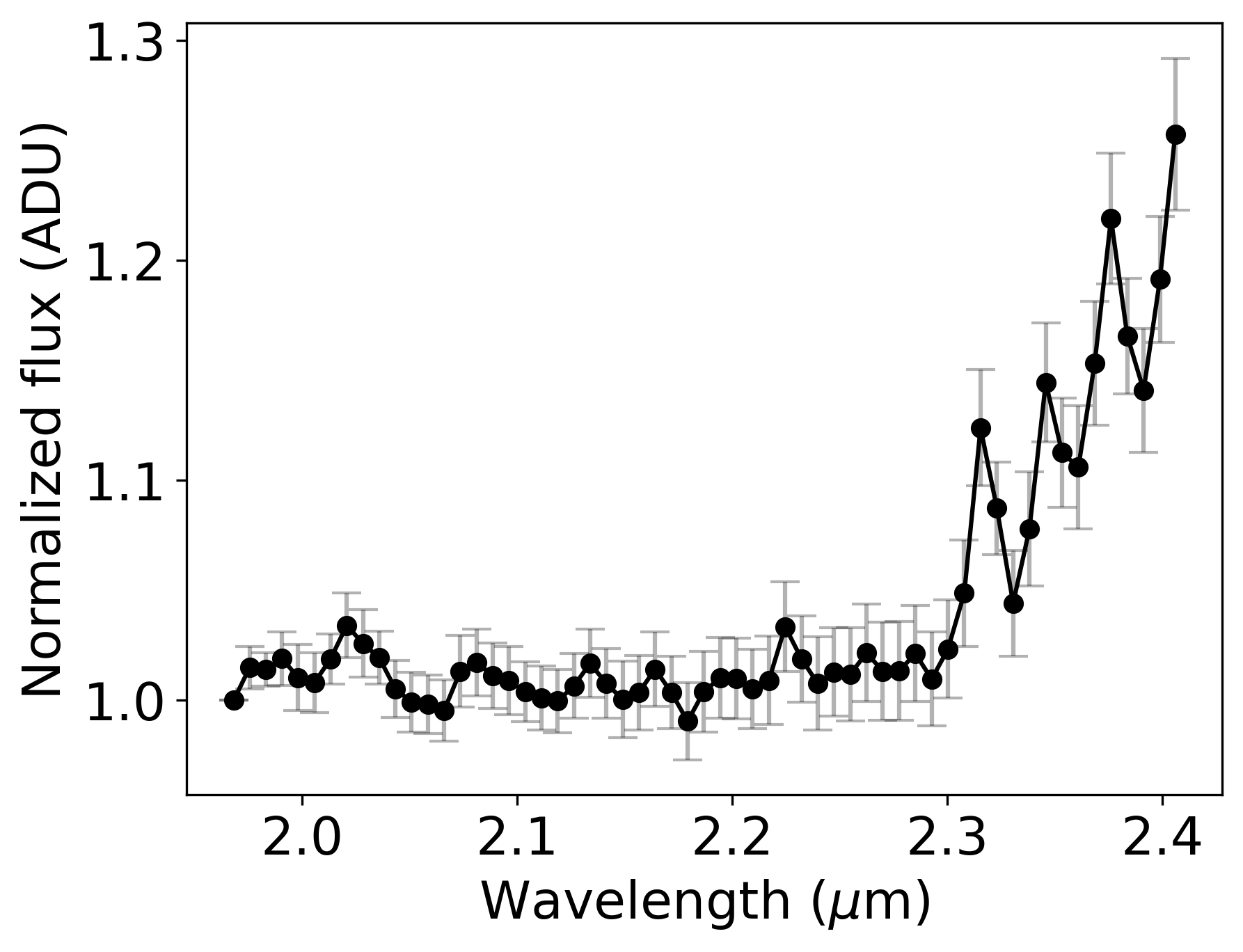}
\caption{The normalized flux as a function of wavelength in the K-band observed from the MYSTIC instrument. The flux at longer wavelengths, $\lambda>2.3~\mu$m, appears to be composed of emission lines that increase the observed flux. 
}
\label{Fig:flux}
\end{figure}

Figure~\ref{Fig:flux} compares the photometry across K-band wavelengths observed with the MYSTIC instrument. The $\rho$~Cas flux measured from six telescopes is averaged, and then it is divided by the flux of the calibrator stars for the spectral calibration. Since MYSTIC does not measure the absolute flux, we have normalized it by taking the mean flux observed at $\lambda=2.0~\mu$m.

\bibliography{main.bib}{}

\begin{thebibliography}{}
\expandafter\ifx\csname natexlab\endcsname\relax\def\natexlab#1{#1}\fi
\providecommand{\url}[1]{\href{#1}{#1}}
\providecommand{\dodoi}[1]{doi:~\href{http://doi.org/#1}{\nolinkurl{#1}}}
\providecommand{\doeprint}[1]{\href{http://ascl.net/#1}{\nolinkurl{http://ascl.net/#1}}}
\providecommand{\doarXiv}[1]{\href{https://arxiv.org/abs/#1}{\nolinkurl{https://arxiv.org/abs/#1}}}

\bibitem[{{Anugu} {et~al.}(2020){Anugu}, {Le Bouquin}, {Monnier}, {Kraus}, {Setterholm}, {Labdon}, {Davies}, {Lanthermann}, {Gardner}, {Ennis}, {Johnson}, {Ten Brummelaar}, {Schaefer}, \& {Sturmann}}]{Anugu2020}
{Anugu}, N., {Le Bouquin}, J.-B., {Monnier}, J.~D., {et~al.} 2020, \aj, 160, 158, \dodoi{10.3847/1538-3881/aba957}

\bibitem[{{Anugu} {et~al.}(2023){Anugu}, {Baron}, {Gies}, {Lanthermann}, {Schaefer}, {Shepard}, {Brummelaar}, {Monnier}, {Kraus}, {Le Bouquin}, {Davies}, {Ennis}, {Gardner}, {Labdon}, {Roettenbacher}, {Setterholm}, {Vollmann}, \& {Sigismondi}}]{Anugu2023}
{Anugu}, N., {Baron}, F., {Gies}, D.~R., {et~al.} 2023, \aj, 166, 78, \dodoi{10.3847/1538-3881/ace59d}

\bibitem[{{Bailer-Jones} {et~al.}(2021){Bailer-Jones}, {Rybizki}, {Fouesneau}, {Demleitner}, \& {Andrae}}]{BailerJones2021}
{Bailer-Jones}, C.~A.~L., {Rybizki}, J., {Fouesneau}, M., {Demleitner}, M., \& {Andrae}, R. 2021, \aj, 161, 147, \dodoi{10.3847/1538-3881/abd806}

\bibitem[{{Bailer-Jones} {et~al.}(2018){Bailer-Jones}, {Rybizki}, {Fouesneau}, {Mantelet}, \& {Andrae}}]{BailerJones2018}
{Bailer-Jones}, C.~A.~L., {Rybizki}, J., {Fouesneau}, M., {Mantelet}, G., \& {Andrae}, R. 2018, \aj, 156, 58, \dodoi{10.3847/1538-3881/aacb21}

\bibitem[{{Baines} {et~al.}(2023){Baines}, {Clark}, {Schmitt}, {Stone}, \& {von Braun}}]{Baines2023}
{Baines}, E.~K., {Clark}, James~H., I., {Schmitt}, H.~R., {Stone}, J.~M., \& {von Braun}, K. 2023, \aj, 166, 268, \dodoi{10.3847/1538-3881/ad08be}

\bibitem[{{Beardsley}(1961)}]{Beardsley1961ApJS....5..381B}
{Beardsley}, W.~R. 1961, \apjs, 5, 381, \dodoi{10.1086/190058}

\bibitem[{{Bl{\"o}cker} {et~al.}(1999){Bl{\"o}cker}, {Balega}, {Hofmann}, {Lichtenth{\"a}ler}, {Osterbart}, \& {Weigelt}}]{Blocker1999A&A...348..805B}
{Bl{\"o}cker}, T., {Balega}, Y., {Hofmann}, K.~H., {et~al.} 1999, \aap, 348, 805, \dodoi{10.48550/arXiv.astro-ph/9906473}

\bibitem[{{Bourg\'{e}s} {et~al.}(2017){Bourg\'{e}s}, {Mella}, {Lafrasse}, {Duvert}, {Chelli}, {Le Bouquin}, {Delfosse}, \& {Chesneau}}]{Bourges2017}
{Bourg\'{e}s}, L., {Mella}, G., {Lafrasse}, S., {et~al.} 2017, VizieR Online Data Catalog, II/346

\bibitem[{{Bowen}(1992)}]{Bowen1992iesh.conf..104B}
{Bowen}, G.~H. 1992, in Instabilities in Evolved Super- and Hypergiants, ed. C.~{de Jager} \& H.~{Nieuwenhuijzen}, 104

\bibitem[{{Boyarchuk} {et~al.}(1988){Boyarchuk}, {Boyarchuk}, \& {Petrov}}]{Boyarchuk1988TarOT..92...40B}
{Boyarchuk}, A.~A., {Boyarchuk}, M.~E., \& {Petrov}, P.~P. 1988, Tartu Astrofuusika Observatoorium Teated, 92, 40

\bibitem[{{Castro-Carrizo} {et~al.}(2007){Castro-Carrizo}, {Quintana-Lacaci}, {Bujarrabal}, {Neri}, \& {Alcolea}}]{Castro-Carrizo2007A&A...465..457C}
{Castro-Carrizo}, A., {Quintana-Lacaci}, G., {Bujarrabal}, V., {Neri}, R., \& {Alcolea}, J. 2007, \aap, 465, 457, \dodoi{10.1051/0004-6361:20066169}

\bibitem[{{Chesneau} {et~al.}(2014){Chesneau}, {Meilland}, {Chapellier}, {Millour}, {van Genderen}, {Naz{\'e}}, {Smith}, {Spang}, {Smoker}, {Dessart}, {Kanaan}, {Bendjoya}, {Feast}, {Groh}, {Lobel}, {Nardetto}, {Otero}, {Oudmaijer}, {Tekola}, {Whitelock}, {Arcos}, {Cur{\'e}}, \& {Vanzi}}]{Chesneau2014A&A...563A..71C}
{Chesneau}, O., {Meilland}, A., {Chapellier}, E., {et~al.} 2014, \aap, 563, A71, \dodoi{10.1051/0004-6361/201322421}

\bibitem[{{Chiavassa} {et~al.}(2022){Chiavassa}, {Kudritzki}, {Davies}, {Freytag}, \& {de Mink}}]{Chiavassa2022}
{Chiavassa}, A., {Kudritzki}, R., {Davies}, B., {Freytag}, B., \& {de Mink}, S.~E. 2022, \aap, 661, L1, \dodoi{10.1051/0004-6361/202243568}

\bibitem[{{de Jager}(1998)}]{deJager1998A&ARv...8..145D}
{de Jager}, C. 1998, \aapr, 8, 145, \dodoi{10.1007/s001590050009}

\bibitem[{{de Jager} {et~al.}(1997){de Jager}, {Lobel}, \& {Israelian}}]{deJager1997A&A...325..714D}
{de Jager}, C., {Lobel}, A., \& {Israelian}, G. 1997, \aap, 325, 714

\bibitem[{{Dorn-Wallenstein} {et~al.}(2023){Dorn-Wallenstein}, {Neugent}, \& {Levesque}}]{Dorn-Wallenstein2023ApJ...959..102D}
{Dorn-Wallenstein}, T.~Z., {Neugent}, K.~F., \& {Levesque}, E.~M. 2023, \apj, 959, 102, \dodoi{10.3847/1538-4357/ad0725}

\bibitem[{{Drout} {et~al.}(2012){Drout}, {Massey}, \& {Meynet}}]{Drout2012ApJ...750...97D}
{Drout}, M.~R., {Massey}, P., \& {Meynet}, G. 2012, \apj, 750, 97, \dodoi{10.1088/0004-637X/750/2/97}

\bibitem[{{Dupree} {et~al.}(2022){Dupree}, {Strassmeier}, {Calderwood}, {Granzer}, {Weber}, {Kravchenko}, {Matthews}, {Montarg{\`e}s}, {Tappin}, \& {Thompson}}]{Dupree2022}
{Dupree}, A.~K., {Strassmeier}, K.~G., {Calderwood}, T., {et~al.} 2022, \apj, 936, 18, \dodoi{10.3847/1538-4357/ac7853}

\bibitem[{{Ekstr{\"o}m} {et~al.}(2012){Ekstr{\"o}m}, {Georgy}, {Eggenberger}, {Meynet}, {Mowlavi}, {Wyttenbach}, {Granada}, {Decressin}, {Hirschi}, {Frischknecht}, {Charbonnel}, \& {Maeder}}]{Ekstrom2012A&A...537A.146E}
{Ekstr{\"o}m}, S., {Georgy}, C., {Eggenberger}, P., {et~al.} 2012, \aap, 537, A146, \dodoi{10.1051/0004-6361/201117751}

\bibitem[{{Georgy}(2012)}]{Georgy2012A&A...538L...8G}
{Georgy}, C. 2012, \aap, 538, L8, \dodoi{10.1051/0004-6361/201118372}

\bibitem[{{Glatzel} \& {Kraus}(2024)}]{Glatzel2024MNRAS.529.4947G}
{Glatzel}, W., \& {Kraus}, M. 2024, \mnras, 529, 4947, \dodoi{10.1093/mnras/stae861}

\bibitem[{{Gorlova} {et~al.}(2006){Gorlova}, {Lobel}, {Burgasser}, {Rieke}, {Ilyin}, \& {Stauffer}}]{Gorlova2006}
{Gorlova}, N., {Lobel}, A., {Burgasser}, A.~J., {et~al.} 2006, \apj, 651, 1130, \dodoi{10.1086/507590}

\bibitem[{{Groh}(2014)}]{Groh2014A&A...572L..11G}
{Groh}, J.~H. 2014, \aap, 572, L11, \dodoi{10.1051/0004-6361/201424852}

\bibitem[{{Haubois} {et~al.}(2019){Haubois}, {Norris}, {Tuthill}, {Pinte}, {Kervella}, {Girard}, {Kostogryz}, {Berdyugina}, {Perrin}, {Lacour}, {Chiavassa}, \& {Ridgway}}]{Haubois2019A&A...628A.101H}
{Haubois}, X., {Norris}, B., {Tuthill}, P.~G., {et~al.} 2019, \aap, 628, A101, \dodoi{10.1051/0004-6361/201833258}

\bibitem[{{Humphreys} \& {Jones}(2022)}]{Humphreys2022}
{Humphreys}, R.~M., \& {Jones}, T.~J. 2022, \aj, 163, 103, \dodoi{10.3847/1538-3881/ac46ff}

\bibitem[{{Humphreys} {et~al.}(2023){Humphreys}, {Jones}, \& {Martin}}]{Humphreys2023AJ....166...50H}
{Humphreys}, R.~M., {Jones}, T.~J., \& {Martin}, J.~C. 2023, \aj, 166, 50, \dodoi{10.3847/1538-3881/acdd6c}

\bibitem[{{Humphreys} {et~al.}(2006){Humphreys}, {Jones}, {Polomski}, {Koppelman}, {Helton}, {McQuinn}, {Gehrz}, {Woodward}, {Wagner}, {Gordon}, {Hinz}, \& {Willner}}]{Humphreys2006AJ....131.2105H}
{Humphreys}, R.~M., {Jones}, T.~J., {Polomski}, E., {et~al.} 2006, \aj, 131, 2105, \dodoi{10.1086/500811}

\bibitem[{{Jones} {et~al.}(2023){Jones}, {Shenoy}, \& {Humphreys}}]{Jones2023}
{Jones}, T.~J., {Shenoy}, D., \& {Humphreys}, R. 2023, Research Notes of the American Astronomical Society, 7, 92, \dodoi{10.3847/2515-5172/acd37f}

\bibitem[{{Jura} \& {Kleinmann}(1990)}]{Jura1990ApJ...351..583J}
{Jura}, M., \& {Kleinmann}, S.~G. 1990, \apj, 351, 583, \dodoi{10.1086/168496}

\bibitem[{{Kilpatrick} {et~al.}(2021){Kilpatrick}, {Drout}, {Auchettl}, {Dimitriadis}, {Foley}, {Jones}, {DeMarchi}, {French}, {Gall}, {Hjorth}, {Jacobson-Gal{\'a}n}, {Margutti}, {Piro}, {Ramirez-Ruiz}, {Rest}, \& {Rojas-Bravo}}]{Kilpatrick2021MNRAS.504.2073K}
{Kilpatrick}, C.~D., {Drout}, M.~R., {Auchettl}, K., {et~al.} 2021, \mnras, 504, 2073, \dodoi{10.1093/mnras/stab838}

\bibitem[{{Kiss} {et~al.}(2006){Kiss}, {Szab{\'o}}, \& {Bedding}}]{Kiss2006MNRAS.372.1721K}
{Kiss}, L.~L., {Szab{\'o}}, G.~M., \& {Bedding}, T.~R. 2006, \mnras, 372, 1721, \dodoi{10.1111/j.1365-2966.2006.10973.x}

\bibitem[{{Klochkova}(2019)}]{Klochkova2019AstBu..74..475K}
{Klochkova}, V.~G. 2019, Astrophysical Bulletin, 74, 475, \dodoi{10.1134/S1990341319040138}

\bibitem[{{Klochkova} {et~al.}(1997){Klochkova}, {Chentsov}, \& {Panchuk}}]{Klochkova1997MNRAS.292...19K}
{Klochkova}, V.~G., {Chentsov}, E.~L., \& {Panchuk}, V.~E. 1997, \mnras, 292, 19, \dodoi{10.1093/mnras/292.1.19}

\bibitem[{{Klochkova} {et~al.}(2014){Klochkova}, {Panchuk}, {Tavolzhanskaya}, \& {Usenko}}]{Klochkova2014}
{Klochkova}, V.~G., {Panchuk}, V.~E., {Tavolzhanskaya}, N.~S., \& {Usenko}, I.~A. 2014, Astronomy Reports, 58, 101, \dodoi{10.1134/S1063772913120044}

\bibitem[{{Klochkova} {et~al.}(2002){Klochkova}, {Yushkin}, {Chentsov}, \& {Panchuk}}]{Klochkova2002ARep...46..139K}
{Klochkova}, V.~G., {Yushkin}, M.~V., {Chentsov}, E.~L., \& {Panchuk}, V.~E. 2002, Astronomy Reports, 46, 139, \dodoi{10.1134/1.1451927}

\bibitem[{{Kourniotis} {et~al.}(2022){Kourniotis}, {Kraus}, {Maryeva}, {Borges Fernandes}, \& {Maravelias}}]{Kourniotis2022MNRAS.511.4360K}
{Kourniotis}, M., {Kraus}, M., {Maryeva}, O., {Borges Fernandes}, M., \& {Maravelias}, G. 2022, \mnras, 511, 4360, \dodoi{10.1093/mnras/stac386}

\bibitem[{{Kraus} {et~al.}(2019){Kraus}, {Kolka}, {Aret}, {Nickeler}, {Maravelias}, {Eenm{\"a}e}, {Lobel}, \& {Klochkova}}]{Kraus2019MNRAS.483.3792K}
{Kraus}, M., {Kolka}, I., {Aret}, A., {et~al.} 2019, \mnras, 483, 3792, \dodoi{10.1093/mnras/sty3375}

\bibitem[{{Kusuno} {et~al.}(2013){Kusuno}, {Asaki}, {Imai}, \& {Oyama}}]{Kusuno2013ApJ...774..107K}
{Kusuno}, K., {Asaki}, Y., {Imai}, H., \& {Oyama}, T. 2013, \apj, 774, 107, \dodoi{10.1088/0004-637X/774/2/107}

\bibitem[{{Lambert} {et~al.}(1981){Lambert}, {Hinkle}, \& {Hall}}]{Lambert1981ApJ...248..638L}
{Lambert}, D.~L., {Hinkle}, K.~H., \& {Hall}, D.~N.~B. 1981, \apj, 248, 638, \dodoi{10.1086/159189}

\bibitem[{Le~Bouquin {et~al.}(2024)Le~Bouquin, Anugu, Davies, Gardner, Ibrahim, \& Monnier}]{le_bouquin_2024_12735292}
Le~Bouquin, J.-B., Anugu, N., Davies, C.~L., {et~al.} 2024, CHARA MIRC-X and MYSTIC Data Reduction Pipeline,  Zenodo, \dodoi{10.5281/zenodo.12735292}

\bibitem[{{Leiker} {et~al.}(1989){Leiker}, {Hoff}, \& {Milton}}]{Leiker1989IBVS.3345....1L}
{Leiker}, P.~S., {Hoff}, D.~B., \& {Milton}, R. 1989, Information Bulletin on Variable Stars, 3345, 1

\bibitem[{{Lobel} {et~al.}(1998){Lobel}, {Israelian}, {de Jager}, {Musaev}, {Parker}, \& {Mavrogiorgou}}]{Lobel1998}
{Lobel}, A., {Israelian}, G., {de Jager}, C., {et~al.} 1998, \aap, 330, 659

\bibitem[{{Lobel} {et~al.}(2003){Lobel}, {Dupree}, {Stefanik}, {Torres}, {Israelian}, {Morrison}, {de Jager}, {Nieuwenhuijzen}, {Ilyin}, \& {Musaev}}]{Lobel2003ApJ...583..923L}
{Lobel}, A., {Dupree}, A.~K., {Stefanik}, R.~P., {et~al.} 2003, \apj, 583, 923, \dodoi{10.1086/345503}

\bibitem[{{Maravelias} \& {Kraus}(2022)}]{Maravelias2022JAVSO..50...49M}
{Maravelias}, G., \& {Kraus}, M. 2022, \jaavso, 50, 49, \dodoi{10.48550/arXiv.2112.13158}

\bibitem[{{Martinez} {et~al.}(2021){Martinez}, {Baron}, {Monnier}, {Roettenbacher}, \& {Parks}}]{Martinez2021}
{Martinez}, A.~O., {Baron}, F.~R., {Monnier}, J.~D., {Roettenbacher}, R.~M., \& {Parks}, J.~R. 2021, \apj, 916, 60, \dodoi{10.3847/1538-4357/ac06a5}

\bibitem[{{Mauerhan} {et~al.}(2015){Mauerhan}, {Van Dyk}, {Graham}, {Zheng}, {Clubb}, {Filippenko}, {Valenti}, {Brown}, {Smith}, {Howell}, \& {Arcavi}}]{Mauerhan2015MNRAS.447.1922M}
{Mauerhan}, J.~C., {Van Dyk}, S.~D., {Graham}, M.~L., {et~al.} 2015, \mnras, 447, 1922, \dodoi{10.1093/mnras/stu2541}

\bibitem[{{Maund} {et~al.}(2011){Maund}, {Fraser}, {Ergon}, {Pastorello}, {Smartt}, {Sollerman}, {Benetti}, {Botticella}, {Bufano}, {Danziger}, {Kotak}, {Magill}, {Stephens}, \& {Valenti}}]{Maund2011ApJ...739L..37M}
{Maund}, J.~R., {Fraser}, M., {Ergon}, M., {et~al.} 2011, \apjl, 739, L37, \dodoi{10.1088/2041-8205/739/2/L37}

\bibitem[{{Mauron} \& {Le Borgne}(1986)}]{Mauron1986A&A...168..217M}
{Mauron}, N., \& {Le Borgne}, J.~F. 1986, \aap, 168, 217

\bibitem[{{M{\'e}rand}(2022)}]{Merand2022}
{M{\'e}rand}, A. 2022, in Society of Photo-Optical Instrumentation Engineers (SPIE) Conference Series, Vol. 12183, Optical and Infrared Interferometry and Imaging VIII, ed. A.~{M{\'e}rand}, S.~{Sallum}, \& J.~{Sanchez-Bermudez}, 121831N, \dodoi{10.1117/12.2626700}

\bibitem[{{Monnier} {et~al.}(2012){Monnier}, {Che}, {Zhao}, {Ekstr{\"o}m}, {Maestro}, {Aufdenberg}, {Baron}, {Georgy}, {Kraus}, {McAlister}, {Pedretti}, {Ridgway}, {Sturmann}, {Sturmann}, {ten Brummelaar}, {Thureau}, {Turner}, \& {Tuthill}}]{Monnier2012ApJ...761L...3M}
{Monnier}, J.~D., {Che}, X., {Zhao}, M., {et~al.} 2012, \apjl, 761, L3, \dodoi{10.1088/2041-8205/761/1/L3}

\bibitem[{{Montarg{\`e}s} {et~al.}(2014){Montarg{\`e}s}, {Kervella}, {Perrin}, {Ohnaka}, {Chiavassa}, {Ridgway}, \& {Lacour}}]{Montarges2014}
{Montarg{\`e}s}, M., {Kervella}, P., {Perrin}, G., {et~al.} 2014, \aap, 572, A17, \dodoi{10.1051/0004-6361/201423538}

\bibitem[{{Montarg{\`e}s} {et~al.}(2021){Montarg{\`e}s}, {Cannon}, {Lagadec}, {de Koter}, {Kervella}, {Sanchez-Bermudez}, {Paladini}, {Cantalloube}, {Decin}, {Scicluna}, {Kravchenko}, {Dupree}, {Ridgway}, {Wittkowski}, {Anugu}, {Norris}, {Rau}, {Perrin}, {Chiavassa}, {Kraus}, {Monnier}, {Millour}, {Le Bouquin}, {Haubois}, {Lopez}, {Stee}, \& {Danchi}}]{Montarges2021}
{Montarg{\`e}s}, M., {Cannon}, E., {Lagadec}, E., {et~al.} 2021, \nat, 594, 365, \dodoi{10.1038/s41586-021-03546-8}

\bibitem[{Mérand {et~al.}(2024)Mérand, Astro-mh, \& Gomes}]{antoine_merand_2024_10889235}
Mérand, A., Astro-mh, \& Gomes, T. 2024, amerand/PMOIRED: cleanup and removing examples, 1.2,  Zenodo, \dodoi{10.5281/zenodo.10889235}

\bibitem[{{Neugent} {et~al.}(2010){Neugent}, {Massey}, {Skiff}, {Drout}, {Meynet}, \& {Olsen}}]{Neugent2010ApJ...719.1784N}
{Neugent}, K.~F., {Massey}, P., {Skiff}, B., {et~al.} 2010, \apj, 719, 1784, \dodoi{10.1088/0004-637X/719/2/1784}

\bibitem[{{Neugent} {et~al.}(2012){Neugent}, {Massey}, {Skiff}, \& {Meynet}}]{Neugent2012ApJ...749..177N}
{Neugent}, K.~F., {Massey}, P., {Skiff}, B., \& {Meynet}, G. 2012, \apj, 749, 177, \dodoi{10.1088/0004-637X/749/2/177}

\bibitem[{{Nieuwenhuijzen} \& {de Jager}(1995)}]{Nieuwenhuijzen1995A&A...302..811N}
{Nieuwenhuijzen}, H., \& {de Jager}, C. 1995, \aap, 302, 811

\bibitem[{{Nieuwenhuijzen} \& {de Jager}(2000)}]{Nieuwenhuijzen2000A&A...353..163N}
---. 2000, \aap, 353, 163

\bibitem[{{Nordgren} {et~al.}(1999){Nordgren}, {Germain}, {Benson}, {Mozurkewich}, {Sudol}, {Elias}, {Hajian}, {White}, {Hutter}, {Johnston}, {Gauss}, {Armstrong}, {Pauls}, \& {Rickard}}]{Nordgren1999}
{Nordgren}, T.~E., {Germain}, M.~E., {Benson}, J.~A., {et~al.} 1999, \aj, 118, 3032, \dodoi{10.1086/301114}

\bibitem[{{Norris} {et~al.}(2021){Norris}, {Baron}, {Monnier}, {Paladini}, {Anderson}, {Martinez}, {Schaefer}, {Che}, {Chiavassa}, {Connelley}, {Farrington}, {Gies}, {Kiss}, {Lester}, {Montarg{\`e}s}, {Neilson}, {Majoinen}, {Pedretti}, {Ridgway}, {Roettenbacher}, {Scott}, {Sturmann}, {Sturmann}, {Thureau}, {Vargas}, \& {ten Brummelaar}}]{Norris2021}
{Norris}, R.~P., {Baron}, F.~R., {Monnier}, J.~D., {et~al.} 2021, \apj, 919, 124, \dodoi{10.3847/1538-4357/ac0c7e}

\bibitem[{{Ohnaka} {et~al.}(2011){Ohnaka}, {Weigelt}, {Millour}, {Hofmann}, {Driebe}, {Schertl}, {Chelli}, {Massi}, {Petrov}, \& {Stee}}]{Ohnaka2011}
{Ohnaka}, K., {Weigelt}, G., {Millour}, F., {et~al.} 2011, \aap, 529, A163, \dodoi{10.1051/0004-6361/201016279}

\bibitem[{{Oudmaijer}(1998)}]{Oudmaijer1998A&AS..129..541O}
{Oudmaijer}, R.~D. 1998, \aaps, 129, 541, \dodoi{10.1051/aas:1998404}

\bibitem[{{Oudmaijer} \& {de Wit}(2013)}]{Oudmaijer2013}
{Oudmaijer}, R.~D., \& {de Wit}, W.~J. 2013, \aap, 551, A69, \dodoi{10.1051/0004-6361/201220185}

\bibitem[{{Oudmaijer} {et~al.}(1996){Oudmaijer}, {Groenewegen}, {Matthews}, {Blommaert}, \& {Sahu}}]{Oudmaijer1996MNRAS.280.1062O}
{Oudmaijer}, R.~D., {Groenewegen}, M.~A.~T., {Matthews}, H.~E., {Blommaert}, J.~A.~D.~L., \& {Sahu}, K.~C. 1996, \mnras, 280, 1062, \dodoi{10.1093/mnras/280.4.1062}

\bibitem[{{Percy} {et~al.}(2000){Percy}, {Kolin}, \& {Henry}}]{Percy2000}
{Percy}, J.~R., {Kolin}, D.~L., \& {Henry}, G.~W. 2000, \pasp, 112, 363, \dodoi{10.1086/316541}

\bibitem[{{Perrin} {et~al.}(2005){Perrin}, {Ridgway}, {Verhoelst}, {Schuller}, {Coud{\'e} du Foresto}, {Traub}, {Millan-Gabet}, \& {Lacasse}}]{Perrin2005}
{Perrin}, G., {Ridgway}, S.~T., {Verhoelst}, T., {et~al.} 2005, \aap, 436, 317, \dodoi{10.1051/0004-6361:20042313}

\bibitem[{{Quintana-Lacaci} {et~al.}(2008){Quintana-Lacaci}, {Bujarrabal}, \& {Castro-Carrizo}}]{Quintana-Lacaci2008A&A...488..203Q}
{Quintana-Lacaci}, G., {Bujarrabal}, V., \& {Castro-Carrizo}, A. 2008, \aap, 488, 203, \dodoi{10.1051/0004-6361:200809559}

\bibitem[{{Roettenbacher} {et~al.}(2016){Roettenbacher}, {Monnier}, {Korhonen}, {Aarnio}, {Baron}, {Che}, {Harmon}, {K{\H{o}}v{\'a}ri}, {Kraus}, {Schaefer}, {Torres}, {Zhao}, {Ten Brummelaar}, {Sturmann}, \& {Sturmann}}]{Roettenbacher2016}
{Roettenbacher}, R.~M., {Monnier}, J.~D., {Korhonen}, H., {et~al.} 2016, \nat, 533, 217, \dodoi{10.1038/nature17444}

\bibitem[{{Roettenbacher} {et~al.}(2017){Roettenbacher}, {Monnier}, {Korhonen}, {Harmon}, {Baron}, {Hackman}, {Henry}, {Schaefer}, {Strassmeier}, {Weber}, \& {ten Brummelaar}}]{Roettenbacher2017}
---. 2017, \apj, 849, 120, \dodoi{10.3847/1538-4357/aa8ef7}

\bibitem[{{Sargent}(1961)}]{Sargent1961ApJ...134..142S}
{Sargent}, W. L.~W. 1961, \apj, 134, 142, \dodoi{10.1086/147136}

\bibitem[{{Sblewski}(2022)}]{Sblewski2022}
{Sblewski}, M. 2022, BAV Magazine Spectroscopy, 12, 1

\bibitem[{{Schaefer} {et~al.}(2020){Schaefer}, {ten Brummelaar}, {Gies}, {Anderson}, {Farrington}, {Golden}, {Jones}, {Klement}, {Majoinen}, {Ridgway}, {Sturmann}, {Sturmann}, {Turner}, {Vargas}, {Webster}, \& {Woods}}]{Schaefer2020}
{Schaefer}, G.~H., {ten Brummelaar}, T.~A., {Gies}, D.~R., {et~al.} 2020, in Society of Photo-Optical Instrumentation Engineers (SPIE) Conference Series, Vol. 11446, Optical and Infrared Interferometry and Imaging VII, ed. P.~G. {Tuthill}, A.~{M{\'e}rand}, \& S.~{Sallum}, 1144605, \dodoi{10.1117/12.2562665}

\bibitem[{{Setterholm} {et~al.}(2022){Setterholm}, {Monnier}, {Le Bouquin}, {Anugu}, {Ennis}, {Flores}, {Gardner}, {Ibrahim}, {Jocou}, {Kraus}, {Lanthermann}, {Schaefer}, \& {ten Brummelaar}}]{Setterholm2022}
{Setterholm}, B.~R., {Monnier}, J.~D., {Le Bouquin}, J.-B., {et~al.} 2022, in Society of Photo-Optical Instrumentation Engineers (SPIE) Conference Series, Vol. 12183, Optical and Infrared Interferometry and Imaging VIII, ed. A.~{M{\'e}rand}, S.~{Sallum}, \& J.~{Sanchez-Bermudez}, 121830B, \dodoi{10.1117/12.2629437}

\bibitem[{{Shenoy}(2016)}]{Shenoy2016}
{Shenoy}, D.~P. 2016, PhD thesis, University of Minnesota

\bibitem[{{Smith}(2014)}]{Nathan2014ARA&A..52..487S}
{Smith}, N. 2014, \araa, 52, 487, \dodoi{10.1146/annurev-astro-081913-040025}

\bibitem[{{Stothers}(1969)}]{Stothers1969ApJ...156..541S}
{Stothers}, R. 1969, \apj, 156, 541, \dodoi{10.1086/149987}

\bibitem[{{Stothers} \& {Leung}(1971)}]{Stothers1971A&A....10..290S}
{Stothers}, R., \& {Leung}, K.~C. 1971, \aap, 10, 290

\bibitem[{{Stothers}(2012)}]{Stothers2012}
{Stothers}, R.~B. 2012, \apj, 751, 151, \dodoi{10.1088/0004-637X/751/2/151}

\bibitem[{{Stothers} \& {Chin}(2001)}]{Stothers2001ApJ...560..934S}
{Stothers}, R.~B., \& {Chin}, C.-w. 2001, \apj, 560, 934, \dodoi{10.1086/322438}

\bibitem[{{ten Brummelaar} {et~al.}(2005){ten Brummelaar}, {McAlister}, {Ridgway}, {Bagnuolo}, {Turner}, {Sturmann}, {Sturmann}, {Berger}, {Ogden}, {Cadman}, {Hartkopf}, {Hopper}, \& {Shure}}]{tenBrummelaar2005}
{ten Brummelaar}, T.~A., {McAlister}, H.~A., {Ridgway}, S.~T., {et~al.} 2005, \apj, 628, 453, \dodoi{10.1086/430729}

\bibitem[{{Tessore} {et~al.}(2017){Tessore}, {L{\`e}bre}, {Morin}, {Mathias}, {Josselin}, \& {Auri{\`e}re}}]{Tessore2017A&A...603A.129T}
{Tessore}, B., {L{\`e}bre}, A., {Morin}, J., {et~al.} 2017, \aap, 603, A129, \dodoi{10.1051/0004-6361/201730473}

\bibitem[{{Tiffany} {et~al.}(2010){Tiffany}, {Humphreys}, {Jones}, \& {Davidson}}]{Tiffany2010AJ....140..339T}
{Tiffany}, C., {Humphreys}, R.~M., {Jones}, T.~J., \& {Davidson}, K. 2010, \aj, 140, 339, \dodoi{10.1088/0004-6256/140/2/339}

\bibitem[{{Touhami}(2012)}]{Touhami2012}
{Touhami}, Y.~N. 2012, PhD thesis, Georgia State University

\bibitem[{{van Belle} {et~al.}(2009){van Belle}, {Creech-Eakman}, \& {Hart}}]{vanBelle2009}
{van Belle}, G.~T., {Creech-Eakman}, M.~J., \& {Hart}, A. 2009, \mnras, 394, 1925, \dodoi{10.1111/j.1365-2966.2008.14146.x}

\bibitem[{{van Genderen} {et~al.}(2019){van Genderen}, {Lobel}, {Nieuwenhuijzen}, {Henry}, {de Jager}, {Blown}, {Di Scala}, \& {van Ballegoij}}]{vanGenderen2019}
{van Genderen}, A.~M., {Lobel}, A., {Nieuwenhuijzen}, H., {et~al.} 2019, \aap, 631, A48, \dodoi{10.1051/0004-6361/201834358}

\bibitem[{{Wittkowski} {et~al.}(2017){Wittkowski}, {Abell{\'a}n}, {Arroyo-Torres}, {Chiavassa}, {Guirado}, {Marcaide}, {Alberdi}, {de Wit}, {Hofmann}, {Meilland}, {Millour}, {Mohamed}, \& {Sanchez-Bermudez}}]{Wittkowski2017}
{Wittkowski}, M., {Abell{\'a}n}, F.~J., {Arroyo-Torres}, B., {et~al.} 2017, \aap, 606, L1, \dodoi{10.1051/0004-6361/201731569}

\bibitem[{{Wood}(1979)}]{Wood1979ApJ...227..220W}
{Wood}, P.~R. 1979, \apj, 227, 220, \dodoi{10.1086/156721}

\bibitem[{{Yamamuro} {et~al.}(2007){Yamamuro}, {Nishimaki}, {Motohara}, {Miyata}, \& {Tanaka}}]{Yamamuro2007PASJ...59..973Y}
{Yamamuro}, T., {Nishimaki}, Y., {Motohara}, K., {Miyata}, T., \& {Tanaka}, M. 2007, \pasj, 59, 973, \dodoi{10.1093/pasj/59.5.973}

\bibitem[{{Zsoldos} \& {Percy}(1991)}]{Zsoldos1991A&A...246..441Z}
{Zsoldos}, E., \& {Percy}, J.~R. 1991, \aap, 246, 441

\end{thebibliography}
\bibliographystyle{aasjournal}

\end{document}